%% file: ska_zeeman.tex
\documentclass[a4paper,11pt]{article}
\usepackage{aaskaiid}

\usepackage{orcidlink}
\usepackage{scalerel}
\usepackage{wrapfig} 
\usepackage[leftcaption]{sidecap}
\usepackage[export]{adjustbox}
\usepackage{graphbox}

\newcommand{\skipthis}[1]{}
\newcommand{\blos}{$B_{\rm LOS}$}
\newcommand{\bposdir}{$\hat{B}_{\rm POS}$}
\newcommand\arcsec{\mbox{$^{\prime\prime}$}}%
\newcommand\farcm{\mbox{$.\mkern-4mu^\prime$}}%
\newcommand\farcs{\mbox{$.\!\!^{\prime\prime}$}}%
\newcommand{\kms}{\hbox{km\,s$^{-1}$}}
\newcommand{\uG}{$\mu$G}
\newcommand{\Msun}{\mbox{M$_{\odot}$}}
\newcommand{\Lsun}{\mbox{L$_{\odot}$}}
\newcommand{\gd}{\mbox{$\mathcal{G}$}}
\newcommand{\kd}{\mbox{$\mathcal{K}$}}
\newcommand{\md}{\mbox{$\mathcal{M}$}}
\newcommand{\la}{\,\rlap{\raise 0.5ex\hbox{$<$}}{\lower 0.5ex\hbox{$\sim$}}\,}
\newcommand{\water}{H$_2$O}
\newcommand{\meth}{CH$_3$OH}

\skipthis{

}

\title{Measuring Magnetic Field Strengths in Galactic Star-forming Regions via the Zeeman Effect with the SKA}
\ShortTitle{SKA Zeeman}

\ShortName{Bourke et al.} 

\author[1]{Tyler L. Bourke\orcidlink{0000-0001-7491-0048}}
\affiliation[1]{SKA Observatory, Jodrell Bank, Lower Withington, SK11 9FT, UK}
\emailAdd{tyler.bourke@skao.int}

\author[2,3]{Tao-Chung Ching\orcidlink{0000-0001-8516-2532}}
\emailAdd{chingtaochung@gmail.com}
\affiliation[2]{National Radio Astronomy Observatory,1011 Lopezville Rd., Socorro, NM 87801, USA}
\affiliation[3]{National Astronomical Observatories, Chinese Academy of Sciences, Beijing 100101, China}
\author[4]{Laura Fissel\orcidlink{0000-0002-4666-609X}}
\affiliation[4]{Department of Physics, Engineering Physics and Astronomy, Queen’s University, 64 Bader Lane, Kingston, Canada, K7L 3N6}
\emailAdd{laura.fissel@queensu.ca}
\author[5]{James A. Green\orcidlink{0000-0002-2670-188X}}
\affiliation[5]{SKAO, ARRC Building, 26 Dick Perry Avenue, Kensington WA 6151, Australia}
\emailAdd{Jimi.Green@skao.int}
\author[6,7]{Jihye Hwang\orcidlink{0000-0001-7866-2686}}
\affiliation[6]{Institute for Advanced Study, Kyushu University, Japan}
\affiliation[7]{Department of Earth and Planetary Sciences, Faculty of Science, Kyushu University, Nishi-ku, Fukuoka 819-0395, Japan}
\emailAdd{astrojhwang@gmail.com}
\author[8,9]{A. M. Jacob\orcidlink{0000-0001-7838-3425}}
\affiliation[8]{ I. Physikalisches Institut, Universit\"at zu K\"oln, Z\"ulpicher Str. 77, D-50937 K\"oln, Germany}
\affiliation[9]{Max-Planck-Institut für Radioastronomie, Auf dem Hügel 69, 53121, Bonn, Germany}
\author[10]{Boy Lankhaar\orcidlink{0000-0001-8975-9926}}
\affiliation[10]{Institute of Theoretical Astrophysics, University of Oslo, P.O. Box 1029, Blindern, 0315 Oslo, Norway}
\author[2]{Emmanuel Momjian\orcidlink{0000-0003-3168-5922}}
\emailAdd{emomjian@nrao.edu}
\author[11]{Kate Pattle\orcidlink{0000-0002-8557-3582}}
\affiliation[11]{Department of Physics and Astronomy, University College London, Gower Street, London WC1E 6BT, UK}
\emailAdd{k.pattle@ucl.ac.uk}
\author[12]{A.~P.~Sarma\orcidlink{0000-0001-6590-552X}}
\affiliation[12]{Department of Physics and Astrophysics, DePaul University, 2219 N. Kenmore Ave., Byrne 211, Chicago, USA}
\emailAdd{asarma@depaul.edu}
\author[13,14]{Mehrnoosh Tahani\orcidlink{0000-0001-8749-1436}}
\affiliation[13]{Department of Physics \& Astronomy, University of South Carolina, Columbia, SC 29208, USA}
\affiliation[14]{Kavli Institute for Particle Astrophysics \& Cosmology (KIPAC), Stanford University, Stanford, CA 94305, USA}
\author[15,16]{Chenoa D. Tremblay\orcidlink{0000-0002-4409-3515}}
\affiliation[15]{SETI Institute, 339 Bernardo Ave, Suite 200, Mountain View, CA 94043, USA}
\emailAdd{ctremblay@seti.org}
\affiliation[16]{Department of Physics and Astronomy, University of New Mexico, Albuquerque, NM 87131, USA}

\author[17]{Timothy Robishaw\orcidlink{0000-0002-4217-5138}}
\emailAdd{tim.robishaw@nrc-cnrc.gc.ca}
\affiliation[17]{Dominion Radio Astrophysical Observatory, Herzberg Astronomy \& Astrophysics Research Centre, National Research Council Canada, 717 White Lake Rd., Kaleden, BC, V0H 1K0, Canada}

\author[18,19]{Tomoya Hirota\orcidlink{0000-0003-1659-095X}}
\emailAdd{tomoya.hirota@nao.ac.jp}
\affiliation[18]{Mizusawa VLBI Observatory, National Astronomical
Observatory of Japan, 2-12 Hoshigaoka-cho, Mizusawa, Oshu-shi, Iwate 023-0861, Japan}
\affiliation[19]{SOKENDAI (The Graduate University for Advanced Studies), 2-21-1
Osawa, Mitaka-shi, Tokyo, 181-8588, Japan}

\author[20]{Kristen L. Thompson\orcidlink{0000-0002-0368-6330}}
\affiliation[20]{Davidson College, Davidson, NC 28035, USA}

\author[21]{Keping Qiu\orcidlink{0000-0002-5093-5088}}
\affiliation[21]{School of Astronomy and Space Science, Nanjing University, 163 Xianlin Avenue,  Nanjing 210023, China}
\emailAdd{kpqiu@nju.edu.cn}

\author[22,1]{Aaryaa Premanand}
\affiliation[22]{London School of Economics and Political Science, Houghton Street, London, WC2A 2AE, UK}
\emailAdd{aaryaapremanand@gmail.com}

\abstract{
Magnetic fields thread the interstellar medium from the largest to the smallest scales and play an important role in molecular cloud evolution and star formation. Quantifying this requires measurements of the field strengths, and the most direct way to measure them is via the Zeeman effect in spectral lines. The effect is subtle for the typical field strengths expected from theory, from a few $\mu$G in diffuse molecular clouds to a few 10s of mG in dense star-forming regions, and detections are scarce. Existing measurements of magnetic field strength suggest dense clouds and cores are marginally supercritical (cannot prevent collapse, but can inhibit it), but may be biased due to small sample sizes.  Zeeman effect measurements tracing different scales and densities within molecular clouds can reveal the variation of field strengths, providing critical measurements to address the question of whether star formation is primarily regulated by magnetic fields or turbulence on different scales.  Observations with SKA precursors such MeerKAT and FAST are beginning to increase the number of Zeeman effect detections in nearby star-forming regions. The SKA will extend their reach to many regions within our Galaxy that are best representative of where most stars form, while zooming in on the densest star-forming regions, providing a statistical basis for the role of magnetic fields in molecular cloud evolution and star formation. We present predictions and plans for Zeeman effect observations with the SKA telescopes, demonstrating the significant advances they will provide for studies of magnetic fields in molecular clouds.
}

\begin{document}
\maketitle


\section{Introduction} 

Stars form in dense cores within filaments in clouds of molecular gas and dust. The contraction and collapse of gas and dust to form stars occurs over molecular cloud size scales of 10s of pc down to the sub-au scales of protostars, and spans many orders of magnitude in density. The evolution of molecular clouds and their ability to form stars is driven primarily by the interplay between gravity, magnetic fields, and turbulence.  Star formation is an inefficient process: the rate of star formation within our Galaxy is much lower than would be the case if it were driven by gravity alone. Molecular clouds convert only a small fraction of their mass into stars during their lifetimes \citep*{ZP74, ZE74, Elmegreen1985}.  The key challenges to understanding the inefficiency of star formation are determining which of the two opposing forces, magnetic fields or turbulence, is most important in regulating star formation on different size scales,  whether that influence is dominant in suppressing star formation or simply slowing it down, and to what degree.  Understanding the reasons why star formation is inefficient are extremely important as the star-formation rate (SFR) has consequences for our understanding of planet formation and galaxy evolution \citep{Evans1999}. 

Understanding the extent of the influence of magnetic fields on star formation has been challenging, due to the difficulty in tracing field structure and measuring field strengths on different size scales within molecular clouds using current instrumentation \citep{crutcher2012, han2017, li2021, pattle2023}.  To address the question of the importance of magnetic fields in regulating star formation, careful measurement of field structure and strength on the different size scales and densities within molecular clouds is required. Additionally, studies have shown that magnetic morphology can reveal structures invisible to total emission observations \citep{Tahanietal2022P, Mohammedetal2024}, the energy budgets in molecular clouds \citep{ Tahanietal2023}, and formation and evolution of clouds \citep{Tahanietal2022O, Tahani2022}. 

Field structure within molecular clouds is best traced via polarised dust emission at mm and submm wavelengths, revealing the plane-of-sky structure \citep{liu2022, pattle2023}.  The number of maps of field structure has increased significantly in recent years, primarily due to large programs on SOFIA and in particular the JCMT \citep[e.g., the BISTRO survey;][]{wt2017, pattle2017}.  Together with Planck polarisation maps, these have increased our knowledge of how field structure varies from large to small scales in different star-forming regions.  Polarisation maps at sub-core scales are now emerging from ALMA observations which will add to our understanding of field variations down to the scale of protostars.   Although field strength can be indirectly inferred from field structure maps via the Davis-Chandrasekhar-Fermi (DCF) method (\S\ref{sec-dust}, these estimates rely on assumptions and there are caveats which may not always be fulfilled \citep{heiles2009, liu2022, pattle2023}.

The most direct measure of magnetic field strength is through observations of the Zeeman effect in spectral lines of selected atoms and molecules at radio and mm wavelengths (see reviews by \cite{heiles1993, crutcher2019}).  For the typical physical conditions found throughout the bulk of a molecular cloud, the magnitude of the Zeeman effect is small, requiring very sensitive observations for its detection (see \S\ref{sec-zee}), and only the value of the line-of-sight (LOS) component of the field can be inferred (the exception is observations of the Zeeman effect in some maser lines of OH, where the full field strength can be recovered - see \S\ref{sec-masers}).   
 
As a result, the number of Zeeman effect detections remains small, even after decades of effort, due to the need for long integration times using radio telescopes with stable and well-understood polarimetric responses.  However, the power of Zeeman measurements in providing the most accurate field strength measurements highlights their importance for mapping the three-dimensional (3D) structure of interstellar magnetic fields (see chapter by \cite{Tahani2026.SKA}) and for determining the energy budgets within molecular clouds.  The few existing measurements of field strength suggest clouds \& cores are marginally supercritical (cannot prevent collapse, but can slow the rate of star formation), but this result may be biased due to small sample sizes.  

\begin{figure}[t!]
    \centering
    \includegraphics[width=\columnwidth]{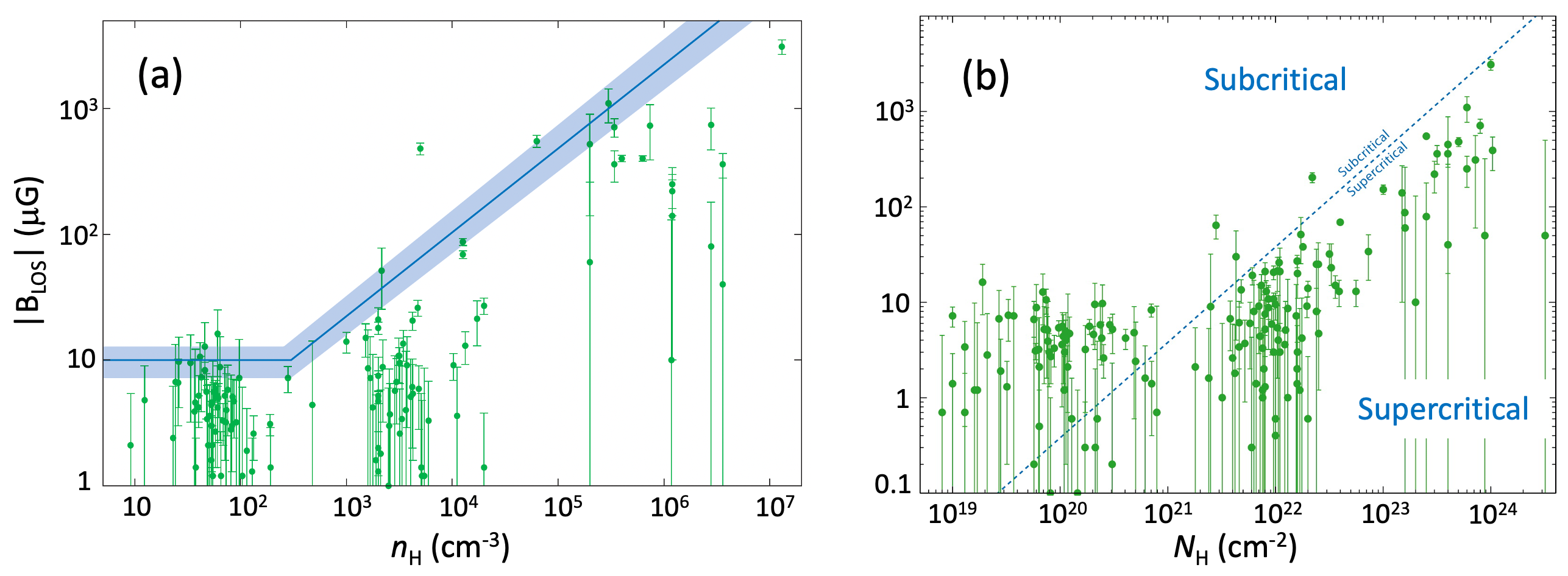}
    \caption{(a) The set of diffuse cloud and molecular cloud Zeeman effect measurements of the magnitude of the line-of-sight component $B_{LOS}$ of the magnetic vector $B$ and their 1$\sigma$ uncertainties, plotted against $n_{\rm H}$ (= $n$(HI) or 2$n$(H$_2$)) for HI and molecular clouds, respectively \citep{crutcher2010}. Although Zeeman effect measurements give the direction of the line-of-sight component as well as the magnitude, only the magnitudes are plotted. The solid blue line shows the most probable maximum values for $B_{\rm TOT}(n_{\rm H})$ determined from the plotted values of $B_{\rm LOS}$ by the Bayesian analysis of \cite{crutcher2010}. Also shown (plotted as light blue shading) are the ranges given by acceptable alternative model parameters to indicate the uncertainty in the model. (b) HI, OH, and CN Zeeman effect measurements of $B_{\rm LOS}$ versus $N_{\rm H} = N_{\rm HI} + 2N_{\rm H_2}$. The dashed blue line is for a
    critical mass-to-flux ratio (M/$\Phi$)$_{\rm crit}$ = 3.8 $\times$ 10$^{-21}N_{\rm H}/B$. Measurements above this line are subcritical, those below are supercritical ({\bf see section \ref{sec-metrics}}). Figure adapted from \cite{crutcher2012}.
}
    \label{fig:relations}
\end{figure}

Figure~\ref{fig:relations} summarizes most of the published Zeeman effect measurements toward molecular clouds (and diffuse HI clouds).   Although these plots were presented over 10 years ago, very little new observational data have been published, and therefore there is almost no new information to add to these figures.  This lack of advancement highlights the difficulty in measuring the Zeeman effect with pre-SKA facilities, and the lack of desire for attempting new measurements with new SKA precursors and pathfinders (e.g., ASKAP, MeerKAT, FAST), until now.  The observations are challenging and more often than not have resulted in non-detections.

Fig.~\ref{fig:relations}(a) presents the Zeeman effect measurement results for a set of atomic and molecular clouds compiled by \cite{crutcher2010}, which represents most of the measurements available at that time for which an estimate of the volume density is also available \citep{crutcher2012}.   This plot shows clustering of data around $n\sim10^2$ cm$^{-3}$ from HI 21-cm observations, around $n\sim10^{3.5}$ cm$^{-3}$ from OH 18-cm observations, and a spread of data at higher values of $n$ from CN 3-mm observations.  Many of the measurements are upper limits (i.e., non-detections), but are included as "detections" for the Bayesian analysis performed by \cite{crutcher2010}, shown by the blue line (indicating the most probable value for the total field strength, $B_{\rm TOT}$) and shading in the figure.  

Fig.~\ref{fig:relations}(b) presents $B_{\rm LOS}$ and $N_{\rm H}$ for five major Zeeman effect surveys for which both values are presented \citep{crutcher2012}.  This figure also plots the critical mass-to-flux ratio (M/$\Phi$)$_{\rm crit}$, indicating the regions in the plot for subcritical (magnetically supported) and supercritical (magnetic field cannot prevent gravitational collapse) values (geometric corrections may move the line vertically; \citealp{bourke2001}).  

The analysis of the data presented in these figures suggests:
\begin{enumerate}
    \item field strength is constant at $\sim$10 $\mu$G for $n_{\rm H}$ < 300 cm$^{-3}$ (or at least does not vary strongly), which may mark the transition from diffuse to self-gravitating clouds,
    \item above $n_{\rm H}$ $\sim$ 300 cm$^{-3}$, field strength increases with density as a power law $B \propto n^{\kappa}$ with index $\kappa\sim$0.65, as expected for spherical clouds contracting (or collapsing) with flux-freezing \citep[][but see, e.g., \citealt{mouschovias2010}, for counter-arguments]{mestel1966}, but significantly different from that expected for clouds contracting via ambipolar diffusion of $\sim$0.5 \citep{mouschovias1999},
    \item most molecular clouds on all size scales are magnetically supercritical, or close to the critical mass-to-flux ratio, suggesting that while magnetic fields cannot prevent collapse, they may slow star formation.
\end{enumerate}

These results are interesting but are inferred from a small number of Zeeman effect measurements overall (including upper limits), and very few actual detections of the Zeeman effect.  The interpretation and analysis is also subject to debate \citep{mouschovias2010, jiang2020, zhao2024}. Filling in these figures with many more Zeeman detections, along with careful estimates of the column and number densities of the gas in which the effect is measured, is key to making progress.  


\section{The Zeeman Effect -- Observational Considerations}
\label{sec-zee}

The Zeeman effect and its application to the measurement of field strengths in the ISM and molecular clouds is described in detail by \cite{heiles1993, crutcher1993, robishaw2008, crutcher2019}; and \cite{robishaw2021}.  In this section we provide a brief description of its direct applicability for observations.  

Observationally, the Zeeman effect most often reveals itself as small frequency shifts, $\Delta\nu_{z}$, in the right and left circularly polarized (RCP and LCP, respectively) components of the spectral line with respect to the frequency in the zero field case, $\nu_{\mathrm{o}}$.
The frequency shift is given by 
\begin{eqnarray}
\Delta\nu_{z} & = & \frac{g\, \mu_B}{h} B \nonumber \\
 & = & \frac{Z}{2} B
\label{eqn-freqshift}
\end{eqnarray}
where $g$ is the transition-specific Land\'e $g$-factor, $\mu_B$ is the Bohr magneton, $h$ is Planck's constant, $B$ the magnetic field strength.  The ``Zeeman splitting factor", $Z = 2g\mu_B/h$, is specific for each spectral transition and is listed in Table~\ref{zee-tab}.   The quantum mechanical calculation of $g$-factors is non-trivial, and accurate values for some transitions are still debated (\S\ref{sec-gfactor}).

The magnetic field is determined from the Stokes $V$ spectrum, 
$V = RCP - LCP$,
whereas Stokes $I = RCP + LCP$.  Under most astrophysical conditions,
$\Delta\nu_{z} \ll \Delta \nu$, where $\Delta \nu$ is the full width at half maximum of the spectral line (the exceptions are some OH masers, see \S\ref{sec-oh}), and so detecting the shift between the RCP and LCP
components due to the the Zeeman effect is difficult, and 
complete information about the magnetic field direction and magnitude is 
not obtainable.  In this situation, the Stokes $V$ spectrum allows only for 
the determination of the line-of-sight component of the field strength, 
\blos\ (= $B\cos\theta$, where $\theta$ is the angle between the direction of the magnetic field and
the line of sight), and its sign (i.e., toward or away from the observer). 
It is then reasonable to approximate $RCP - LCP$ by the derivative of $I$, so that 
\begin{eqnarray}
V & = & \Delta\nu_{z} \cos \theta \frac{dI}{d\nu} + \beta I \\
 & = & \frac{Z}{2} B_{LOS} \frac{dI}{d\nu} + \beta I
\label{eqn-zee2}
\end{eqnarray}
where $\beta I$ represents a gain term which introduces a replica of 
the $I$ spectrum scaled by the factor $\beta$ into the $V$ spectrum.
This gain term is required since in a real experimental setup there will be polarisation leakage due to imperfections within the signal paths and the telescope feeds (linear or circular) \citep{cotton1999,bhatnagar2001,thompson2017}.  Careful calibration of the data following well established polarisation calibration procedures can mitigate this issue so that the scaling factor $\beta$ is $\ll$ 1.
The Zeeman effect reveals itself in the $V$ spectrum in the small
splitting approximation as a characteristic sideways ``S", or ``Zeeman pattern" (essentially a derivative of a Gaussian line profile).  

An advantage of observing the Zeeman effect at SKA frequencies is that the 
{\em frequency offset} due to the Zeeman
effect is {\em independent} of the line frequency, whereas the {\em Doppler
broadened line width} is {\em proportional} to the line frequency.  The
ratio of the Zeeman effect to the line width decreases as the frequency
of the line increases, and so low frequency lines are preferred.


\section{Zeeman Effect Tracers at Frequencies Covered by the SKA Telescopes}

\begin{table}[!ht]
	{\small 
    \begin{center}
	\caption{Zeeman effect tracers at SKA frequencies of relevance to star formation$^a$}
	\label{zee-tab}
	\begin{tabular}{llccc} 
		\hline
	Species  & Transition & Frequency & Splitting factor ($Z^b$) & R.I.$^c$\\
      &  & (GHz) & (Hz/$\mu$G) &   \\
		\hline
		HI & $^2S_{1/2}\ F = 1-0$ & 1.420406 & 2.8 & 1  \\
        H/He/C & RRLs & 0.05-15.4 & 2.8 & 1  \\
        CH $^2\Pi_{3/2}$ & $J=3/2, F=2-2$ & 0.701677 & 1.81 & 9 \\
         & $J=3/2, F=1-1$ & 0.724788$^d$ & 3.03 & 5 \\
         & $J=5/2, F=3-3$ & 4.847768 & 1.15 & 10 \\
         & $J=5/2, F=2-2$ & 4.870059 & 1.62 & 7 \\
         OH $^2\Pi_{3/2}$ & $J=3/2, F=1-1$ & 1.665402 & 3.27 & 5 \\
         & $J=3/2, F=2-2$ & 1.667359 & 1.96 & 9 \\
         & $J=3/2, F=2-1$ & 1.720530$^e$ & 1.31 & 1 \\
         & $J=5/2, F=2-2$ & 6.030748$^e$ & 1.59 & 7 \\
         & $J=5/2, F=3-3$ & 6.035093$^e$ & 1.13 & 10 \\
         & $J=7/2, F=3-3$ & 13.434637 & 1.03 & 10 \\
         & $J=7/2, F=4-4$ & 13.441417$^e$ & 0.80 & 13 \\
         C$_4$H & $N=1-0, J=3/2-1/2, F=1-0$  & 9.49306 & 2.27 & 2.5 \\
         & $N=1-0, J=3/2-1/2, F=2-1$  & 9.49762 & 0.7 & 7.4 \\
         & $N=1-0, J=3/2-1/2, F=1-1$  & 9.50800 & 2.53 & 1.9 \\
         & $\mathit{N=2-1, J=5/2-3/2, F=2-1}$ & $\mathit{19.01472}$ & $\mathit{0.80}$ & $\mathit{7.8}$ \\
         & $\mathit{N=2-1, J=3/2-1/2, F=3-2}$ & $\mathit{19.01514}$ & $\mathit{0.47}$ & $\mathit{12.3}$ \\
         CCS & $J_N = 1_0-0_1$ & 11.119446 & 0.81 & 1 \\
             & $\mathit{J_N = 2_1-1_2}$ & $\mathit{22.344033}$ & $\mathit{0.77}$ & $\mathit{1}$ \\
        SO & $J_N = 1_2-1_1$ & 13.04381 & 1.93 & 1 \\
        \hline
        \vspace*{-6pt} \\
        OH $^2\Pi_{1/2}$ & $J=1/2, F=1-0$  & 4.765562$^e$ & -0.0033 & 1 \\
        CH$_3$OH & $5_1-6_0$ A$^+$  & 6.668519$^e$ & -0.00114 & 1 \\
        & $2_0-3_{-1}$ E & 12.17859$^e$ & -0.001 & 1 \\
        H$_2$O & $\mathit{6_{16}-5_{23}}\;\ F=7-6$ & $\mathit{22.235080^e}$ & $\mathit{0.0021}$ & $\mathit{1}$ \\
        \hline
	\end{tabular}\\
    \end{center}
    \vspace{3pt}
    Table Notes: $^a$Table is adapted from \cite{heiles1993, robishaw2008, crutcher2019} and updated; $^b$All values of $Z$ have been re-calculated, and previously published incorrect values have been updated (e.g., \cite{lankhaar2018}; Lankhaar, in preparation), except H$_2$O \citep{fiebig1989, nedoluha1992, sarma2002}; $^c$R.I. = Relative Intensity: the relative sensitivity to the Zeeman effect for lines from a single species arising from common levels can be estimated by comparing the product of $Z\; \times$ R.I.; $^d$The frequency for the CH $J=3/2, F=1-1$ transition is corrected from that given in \cite{crutcher2019}; $^e$typically seen as a maser.

    }
\end{table}

As discussed above, the best Zeeman-sensitive thermal-line transitions for tracing magnetic fields typical of those in the ISM and star-forming regions are paramagnetic species with large Land\'e $g$-factors that occur at cm wavelengths. Those covered by SKA frequencies and potential upgrades are listed in Table~\ref{zee-tab}. The SKA telescopes will initially cover 50 MHz to 15.4 GHz, while a potential expansion in frequency coverage to $\sim$26 GHz or slightly higher is entirely feasible (e.g., see \href{https://www.skao.int/sites/default/files/documents/d38-ScienceCase_band6_Feb2020.pdf}{SKAO Memo 20-01 "SKA1 Beyond 15 GHz: The Science case for Band 6"}). Strong diamagnetic masers such as CH$_3$OH and H$_2$O with small splitting factors are suitable for regions of high magnetic field strength.  From Table~\ref{zee-tab}, the Zeeman effect in thermal lines has been definitively detected in the OH lines at 1665 \& 1667 MHz, and HI at 1.4 GHz.  Claims of detections in the OH 13 GHz \citep{guesten1994} and CCS 22 GHz lines \citep{koley2022}
exist in the literature but these remain tentative. 


\subsection{Hydroxyl: OH}
\label{sec-oh}

The first detection of the Zeeman effect in a thermal molecular transition was with OH at 1665/1667 MHz in absorption against the HII region NGC 2024 
\citep{ck1983}.  The large Zeeman splitting factors for these transitions, and their ubiquity in absorption and emission at densities typical of molecular clouds ($10^2-10^4$ cm$^{-3}$), means they have been the prime tool for surveys of the Zeeman effect in these regions \citep{crutcher1999b, bourke2001, thompson2019, troland2008, crutcher1993}, and contribute most of the data shown in Fig~\ref{fig:relations}.  Additionally, they constitute all of the secure interferometric molecular Zeeman detections to date \citep[e.g.,][]{roberts1995, crutcher1999a, sarma2000, brogan2001, sarma2013, koley2021}.

The OH 1665/1667 MHz pair provides two advantages for Zeeman effect measurements.  First, they lie close in frequency so they may be observed simultaneously, and second, the ratio of their Zeeman $b$ factors is similar to the inverse of their relative strengths, and so they are approximately equally sensitive to the Zeeman effect.  Conversely, it is difficult to separate $N$(OH) and $T_{\rm ex}$ from observations, unless Local Thermodynamic Equilibrium (LTE) is assumed and the beam filling factor is known.
Values of $N$(H) inferred from $N$(OH) in Fig~\ref{fig:relations} are thus uncertain due to the poorly constrained values of $T_{\rm ex}$ \citep{ebisawa2019, ebisawa2020, harju2000}. 

The excited transitions of OH (4.7 GHz, 6.0 GHz, 13 GHz) listed in Table~\ref{zee-tab} generally exhibit maser activity (\S\ref{sec-masers}), whereas thermal absorption is only occasionally weakly seen.  The single exception is the absorption toward W3(OH) in the 13.434 GHz line, which shows a clear Zeeman effect detection with an inferred field strength of +3.1$\pm$0.4 mG \citep{guesten1994}.  More examples of absorption in this line are needed to trace the field in the high density shocked gas around protostars.


\subsection{Methylidene: CH}

CH has been a promising candidate for Zeeman observations for many years \citep{heiles1993}, but this remains unfulfilled due to the lack of detections of the most promising CH line that can showcase Zeeman splitting: the first rotationally excited state ($^{2}\Pi_{3/2}, N=1, J=3/2$) transitions at 701 and 724\,MHz (Table~\ref{zee-tab}). 
To date only weak detections of the CH lines in total intensity using significant observing time have been published.
\cite{Ziurys1985} observed the 700~MHz lines with the Arecibo 305\,m telescope toward the HII region W51A with over four days of observations, with follow-up observations using the NRAO 300\,ft Green Bank telescope. These observations detected the lines in absorption with intensities < 1 K. One or both of these lines have also been detected toward W51M, W3, W43, and Orion B \citep{Turner1988, Ziurys1985}, while the W51A detection has been replicated in at 2.5h observations with the GBT (Tremblay et al.\ in prep.).  More recently, \citet{Tremblay2020} used ASKAP to search for the 700 MHz lines towared RCW 38 with negative results, while \citet{Jacob2024} detected them with the uGMRT toward W51, and modelled them alongside the 3.3~GHz ground-state and 560~$\mu$m excited state lines using detailed non-LTE radiative transfer calculations. These models indicate that the 700~MHz lines trace relatively high-density gas, around 10$^{5}$~cm$^{-3}$, as predicted \citep{heiles1993}.  

The more commonly detected 3.3 GHz transition is not sensitive to the Zeeman effect for practical purposes, due to small Zeeman splitting factors.  While some higher frequency excited state CH transitions do have Zeeman splitting factors of order unity (Lankhaar priv. comm.), they have not yet been detected \citep[e.g.,][]{matthews1986}. Recent detailed modelling suggests they will not be detectable under typical conditions within molecular clouds or the ISM \citep{Jacob2024}.

Although rarely observed, the 700 MHz lines are essential for interpreting the CH pumping cycle, as they directly connect the first excited state to the ground-state levels involved in pumping the 560~$\mu$m line \citep{Jacob2024}. Coupled with the well-understood 3.3~GHz maser, they offer a unique diagnostic of both CH excitation and magnetic fields in regions of high-density molecular gas, a regime currently inaccessible through most other tracers used in the Crutcher relationship \citep{crutcher2012}. With the sensitivity offered by future SKA observations, these lines have the potential to provide the first systematic measurements of magnetic fields in envelopes of massive star-forming regions.


\subsection{Thioxoethenylidene: CCS}

CCS is an early-time chemical tracer, before star formation occurs \citep[e.g.,][]{suzuki1992, aikawa2001, seo2019, chen2025}, tracing densities of a few $\times 10^4$ cm$^{-3}$ in both low- and high-mass star-forming regions.  Most observations have been made with its 22 GHz and 45 GHz lines \citep[e.g.,][]{hirota2009}.   The CCS lines at 11, 22, 33, and 45 GHz all have Zeeman splitting factors $\sim$1 Hz/$\mu$G, and a handful of Zeeman observations have been made, resulting in both tentative and non-detections \citep{shinnaga1999, levin2001, nakamura2019, koley2021}.  A claim of a Zeeman detection in the 11 GHz line toward the prestellar core TMC-1C exists \citep{guesten1990}, inferring a field strength of $\sim$110 $\mu$G with an uncertain $g$-factor.  The line intensity of $\sim$2 K measured with the Effelsberg 100-m is encouraging, but almost no other observations of this line have been published \citep{uchida2001}, and so its utility as a Zeeman tracer remains unclear.  The recent GOTHAM survey of the low-mass prestellar core TMC-1 with the GBT shows the strength of the 11 and 22 GHz lines to be similar, at a few K, which is not only encouraging, but surprising, as under LTE conditions the 22 GHz line should be 2--3 times brighter \citep{xue2025}.  In TMC-1 at least the population of CCS transitions deviate significantly from a Boltzmann distribution in a direction that is favourable for the 11 GHz line -- there are now many observations of the 22 GHz line as part of large K-band surveys of the ammonia lines (NH$_3$) using the GBT, and the Nobeyama 45-m, from which the line brightness of the 11 GHz may be inferred \citep{kaifu2004, hirota2009, hirota2011, seo2019, pineda2025}. Previous estimates of the utility of the 11 GHz line for SKA-Mid Zeeman observations may be overly pessimistic (see \S3.10.2 in \href{https://www.skao.int/sites/default/files/documents/d38-ScienceCase_band6_Feb2020.pdf}{SKAO Memo 20-01 "SKA1 Beyond 15 GHz: The Science case for Band 6"}), but more observations of the 11 GHz line are needed (section \ref{sec:case3}).


\subsection{Other Molecular Zeeman Effect Candidates}

Within the frequency range of the SKA telescopes are two other potential Zeeman candidates, the paramagnetic molecules C$_4$H and SO.  There are a number of C$_4$H transitions around 9.5 and 19 GHz that are sensitive to the Zeeman effect, while SO has a single transition at 13 GHz (Table~\ref{zee-tab}).  The 3 hyperfine components of the $N=1-0, J=3/2-1/2$ line of C$_4$H near 9.5 GHz have an almost equal sensitivity to the Zeeman effect in LTE (the product of their relative intensities and their splitting factors), although the line at 9.497 GHz has a significantly higher intensity.   With three lines of almost equal sensitivity to the Zeeman effect, any possible detection of the effect can be confirmed through obtaining the same result in all three components, as shown for the example of the CN 1--0 Zeeman detections at 3-mm \citep{crutcher1996, crutcher1999c}, and the OH lines at 1.6 GHz.  Further, the three hyperfine components of the $N=1-0, J=1/2-1/2$ line near 9.55 GHz all have Zeeman splitting factors $\sim$1 Hz/$\mu$G, and the 9.551 GHz line has a sensitivity to the Zeeman effect that is similar to that of the 9.5 GHz lines, which would provide further support for any Zeeman effect detection in C$_4$H.

Unfortunately there are very few observations or detections of any of the 9.5 GHz lines \citep{turner2006}.  
For example, the GBT spectral line survey of TMC-1 \citep{xue2025} detected all six of the 9.5 GHz hyperfine lines in their expected LTE ratios, with a peak of $\sim$0.75~K in the 9.497 GHz line. TMC-1 is known to be carbon-rich, so this result may not be typical. A survey of the 9.5 GHz C$_4$H lines toward molecular cloud cores and protostars, particularly those that are known to be carbon-rich, is needed to assess their practicality as Zeeman effect tracers.  

With an extension to the SKA-Mid frequency coverage, the 
C$_4$H lines around 19 GHz become interesting as tracers of the Zeeman effect.  As for the 9.5 GHz lines, there are very few observations of these lines to determine their practicality as Zeeman effect tracers \citep{kaifu2004, gupta2009, xue2025, lis2025}, and so surveys are also required at these frequencies.  In Table~\ref{zee-tab} we provided updated values of the Zeeman-splitting factors for the most promising C$_4$H transitions. 

Sulphur monoxide (SO) has been identified as a Zeeman effect candidate for many years (e.g., Clark \& Johnson 1974, although their interpretation of the broad SO linewidths in Orion as due to Zeeman broadening was incorrect), and most observations of SO in total intensity have focused on the lines $\geq$ 30 GHz.   Calculations of the SO Zeeman-splitting factors have  primarily been for the transitions $\geq$ 30 GHz \citep{bel1989, shinnaga2000, chiong2003, cazzoli2017}, while published attempts to observe the Zeeman effect at these frequencies only exist for the 30 GHz line toward DR21(OH) and Orion B \citep{chiong2003}, with an unconfirmed detection toward DR21(0H) of $-2.1 \pm 1.2$ mG (i.e., less than a 2$\sigma$ result).  

The SO 13 GHz line has been covered in the deep spectral line surveys of TMC-1 \citep{kaifu2004, xue2025}, but not detected (although the 30 GHz line was detected).  As noted above, TMC-1 is carbon-rich, is a starless core, and has a moderate central density.  Chemical models predict that SO has a slow chemical evolution and is more likely to be present in evolved sources (protostars) of higher density, a result that seems to be confirmed by observations \citep{rydbeck1980, codella1997, zinchenko2018}, as well as enhanced in shocks (such as those seen in protostellar outflows, \citealt{pineau1993}), but depleted in HII-only sources \citep{li2015}.  So TMC-1 may not be a good case study for the presence of SO.  \cite{uchida2001} undertook a survey of the SO 11 GHz line with the Effelsberg 100-m of approximately 50 positions (using short integrations) to find candidates for Zeeman observations, detecting about half of them.  Their observations of the Zeeman effect toward 8 sources resulted in no detections.   Notably, all their Zeeman effect targets were protostellar sources, supporting the expectation that high-density protostellar regions (including shocks) may be the best for SO detections.   Observations of SO at mm/submm wavelengths further support this view \citep{tang2024}.  As is the case for C$_4$H, larger surveys of the 13 GHz SO line are needed to assess its practicality as a Zeeman tracer.


\subsection{HI}
\label{sec-HI}

The HI 21-cm line was the first spectral line in which the Zeeman effect was detected in the interstellar medium, providing the earliest direct measurements of magnetic fields in the cold neutral medium (CNM) \citep{Verschuur1968}, from which molecular clouds form. 
One major advantage of HI Zeeman observations is that the 21-cm line is ubiquitous across the Galaxy, and hence HI Zeeman observations remain a fundamental tool for tracing magnetic fields in the ISM, particularly the diffuse, atomic phases.  In molecular clouds HI Zeeman effect observations are possible in absorption lines seen either against a bright continuum region, such as a HII region, or as self-absorption in cold HI.  

In molecular clouds with high-mass star formation, Zeeman effect observations of HI absorption have been made toward 
photodissociation regions (PDRs) in front of compact HII regions  \citep{1993Roberts,1996Crutcher,1999Brogan,2001Brogan,2016Troland}, where the bright background continuum emission often results in deep HI absorption lines and enhances the sensitivity of HI to the Zeeman effect. These PDRs are the thin layer between the ionised region around the massive star(s) and the molecular cloud, and thus can trace densities in excess of $10^4$ cm$^{-3}$. These observations typical measure magnetic field strengths of 10s to a few hunderd \uG.

\begin{SCfigure}[][ht]
\centering
\includegraphics[align=m,width=0.5\columnwidth]{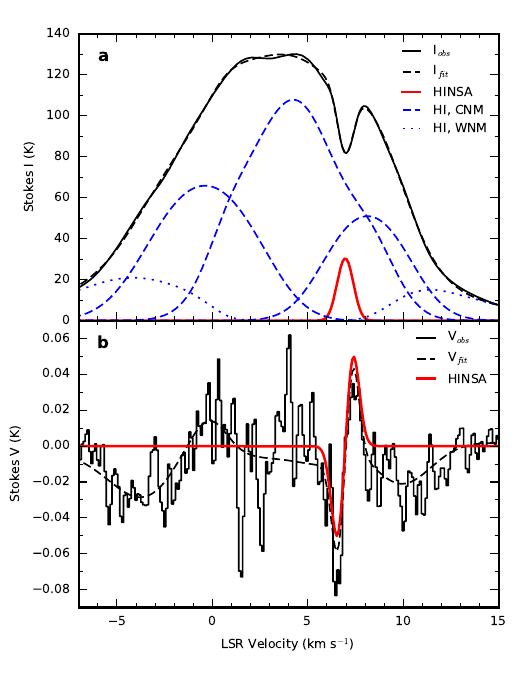}
   \vspace*{-5mm} 
    \caption{
    The Stokes $I$ and $V(v)$ spectra of HI 21-cm line toward L1544 \citep[adapted from][]{2022Ching}. (a) The black profile represents the $I$ spectrum. The red profile represents the absorption from the foreground HINSA component. The blue dashed and dotted profiles represent the emission of the CNM and WNM components, respectively. The black dashed profile represents the sum of the absorption and emission profiles. (b) The black profile represents the $V$ spectrum. The black dashed profile represents the sum of the Zeeman splitting profiles of the five components. The red profile represents the Zeeman splitting profile with $B_{\rm LOS} = +3.8$ $\mu$G of the HINSA component.\\
    \\
    \\} 
    \label{fig:hinsa}   
\end{SCfigure}

In cold molecular clouds, HI narrow self-absorption (HINSA) has been observed, and the HI and OH survey of dark clouds by \citet{2003LG} established a correlation between HINSA and the moderate density gas traced by OH ($\sim$10$^3$ cm$^{-3}$).
Thus HINSA provides another probe of magnetic fields in molecular clouds without high-mass star formation.
The advantages of using HINSA as a Zeeman tracer include: (1) the Land\'e $g$-factor of HINSA is larger than those of most molecules; (2) the intensity of HINSA is usually stronger than other molecular lines; (3)
close to the steady state between H$_2$ formation and destruction, the number density of HINSA is independent of the gas density, making HINSA a promising probe for tracing magnetic fields inside the dense depletion zone that is inaccessible to other Zeeman tracers.
The first HINSA Zeeman effect detection was obtained using the Five-hundred-meter Aperture Spherical radio Telescope (FAST) at 2\farcm9 resolution toward the prestellar core L1544 \citep{2022Ching}. The HINSA Zeeman effect gives $B_{\rm LOS} = +3.8 \pm 0.3$ $\mu$G (Figure \ref{fig:hinsa}), comparable to the strengths derived via the Zeeman measurements of HI absorption toward the surrounding CNM gas of the prestellar core L1544, indicating an early transition from a magnetically subcritical CNM to a supercritical core.

For Zeeman observations of HI absorption against compact HII regions and HINSA toward cold dense cores, the high angular resolution of SKA-Mid will allow for precise mapping of magnetic field strengths in these dense regions.  These observations may be complementary to those using other tracers, as described in this chapter, or they may be the only probe when the other tracers are not present or too weak.  

\subsection{Radio Recombination Lines}

Radio recombination lines (RRLs) of hydrogen, helium and carbon are potential Zeeman tracers in ionised regions, such as PDRs that trace the interface between HII regions and the surrounding molecular cloud, radio jets ejected by protostars, and the surfaces of proto-planetary disks located near to and being ionised by OB stars (e.g., the Orion proplyds).  The magnitude of the Zeeman splitting in RRLs was calculated by \cite{greve1980}, who find that for high $n$ lines ($>30$) of H, He, and C, the result is the same for $n\alpha$ and $n\beta$ transitions for all species: they all have Land\'e $g$-factors = 1 and hence Zeeman-splitting factors of $b = 2.8$ Hz/$\mu$G, the same as for neutral hydrogen.  Many RRLs can be observed simultaneously with SKA-Mid Band 2 (50 lines of H$\alpha$/He$\alpha$/C$\alpha$) and Band 5 (25 lines), while SKA-Low covers over 200 transitions, although they may be better suited to lower density media than found in molecular clouds \citep{thompson2015, oonk2015}.

To date there are no published observations of the Zeeman effect in RRLs at cm wavelengths, perhaps due to their general faintness, and observations of many sources of any type will require the sensitivity of the SKA telescopes \citep{silverglate1984, balser2016}.  The detection of the Zeeman effect at mm wavelengths in the H30$\alpha$ line from the disk of the (presumably) young massive star MWC 349, implying a magnetic field strength of $\sim$22 mG \citep{thum1999}, suggests the effect should be observable at radio frequencies in other sources with observations of sufficient sensitivity.


\subsection{Masers}
\label{sec-masers}

Masers, being compact and bright sources, offer the opportunity to measure in situ magnetic fields at high resolution using the Zeeman effect. There is a significant body of literature on magnetic fields in star-forming regions measured via the Zeeman effect in OH, \water, and \meth\ masers. Both \water\ and \meth\ masers occur in the initial phase of star-forming regions, so they can trace the magnetic field at the very early stages of star formation. It has often been pointed out that masers form in special conditions and therefore the magnetic fields they trace are not representative of the larger environments around them. However, it has been demonstrated that information on the larger-scale magnetic field can be recovered \citep[e.g.,][]{green2012, momjian2012, goddi2017, Robishaw2026.SKA}

As a paramagnetic molecule with a large splitting factor, the Zeeman splitting in OH masers can be much greater than the linewidth, such that the OH maser is the only case in which the total field strength can be measured. In all other instances, being maser or thermal lines, only the line-of-sight component (\blos) can be measured using the Zeeman effect because the width of the Zeeman splitting is much narrower than the width of the line itself. Observations of the Zeeman effect in OH masers have revealed fields of the order of several mG, e.g., 4~mG in Orion KL at 1612~MHz \citep{hansen1982}, 6--10~mG in the compact HII region G49.5$-$0.4(e2) at 1720~MHz \citep{benson1984}, 4.7~mG in W3(OH) at 1665~MHz \citep{harvey1974} and 0.6--21~mG in 18 Galactic massive star-forming regions at 1665 and 1667~MHz \citep{fish2005}. In the excited state transitions, \citet{moran1978} mapped 6.035~GHz maser emission toward W3(OH) at 0\farcs01 resolution and measured magnetic field strengths ranging from 2 to 9 mG, and fields of 0.2--11~mG were found in 30 regions at 6030 and 6035~MHz by \citet{green2015}. Recently, \citet{smits2025} reported a 100~mG field for the 4.7~GHz transition (but their field may be overestimated due to the uncertain value of the Zeeman splitting factor). Another OH transition of interest that will fall within the SKA bands is the 13.4~GHz OH maser, for which \citet{baudry1998} reported a Zeeman detection in W3(OH) (see also \S\ref{sec-oh}).

The Zeeman effect in the 22~GHz \water\ maser transition was first observed by \citet{fiebig1989} with the 100-m Effelsberg Telescope. Interferometric observations of the Zeeman effect in \water\ masers toward the star-forming region W3~IRS~5 were reported by \citet{sarma2001} with the VLBA, and were detected for the first time in circumstellar \water\ masers by \citet{vlemmings2001} with the VLA. Additional observations of 22~GHz \water\ masers carried out over the years (e.g., \citealt{sarma2002}; \citealt{vlemmings2005}; \citealt{alves2012}; \citealt{goddi2017}) have revealed magnetic fields in the range of tens to hundreds of mG in a range of star-forming environments.

Class~I methanol (\meth) masers are collisionally pumped in outflows in star-forming regions whereas Class~II \meth\ masers are radiatively pumped near the high-mass protostars.  Several observations over the years have established that magnetic fields of the order of mG to tens of mG are traced by Class~I \meth\ masers (see, e.g., \citealt{sarma2009}; \citealt{sarma2011}; \citealt{momjian2017}; \citealt{momjian2019}; \citealt{sarma2020}) and Class~II \meth\ masers (see, e.g., \citealt{vlemmings2008};   \citealt{surcis2009}; \citealt{vlemmings2011}; \citet{surcis2015}; \citealt{surcis2022}). 

Several methanol maser transitions (e.g., the bright Class II 6.7 and 12.2 GHz masers) are observable with SKA-Mid, while the sensitivity of SKA-Mid will allow measuring magnetic fields using hitherto unexplored maser lines, such as the Class I methanol maser at 9.9 GHz. 
Overall, maser Zeeman splitting provides insight to the in situ magnetic fields across a range of evolutionary phases of star formation, and a natural synergy to thermal-line Zeeman measurements 
(see chapters by \cite{Robishaw2026.SKA} and \cite{Rygl2026.SKA}) 


\subsection{The Reliability of Zeeman-Splitting Factors}
\label{sec-gfactor}

Often a major uncertainty in measuring field strengths via the Zeeman effect is the availability of reliable Land\'{e} $g$-factors, and hence of Zeeman-splitting factors (Table~\ref{zee-tab}).
The Zeeman effect is due to the coupling of the molecular/atomic magnetic moment to the magnetic field. The principle magnetic moment that generates the Zeeman effect can either be due to electrons, as in the case of radical species, or nuclei, in other cases. The former are called paramagnetic species, and are associated with Zeeman splittings on the order of a Bohr magneton ($\mu_B/h = 1.4\ \mathrm{Hz/\mu G}$), while the latter are called diamagnetic species, and have Zeeman splittings on the order of the nuclear magneton ($\mu_N / h = 0.76\ \mathrm{Hz/ mG}$, i.e., $\sim$ 1500 smaller). The exact scaling of the Zeeman effect with the respective magneton is particular to the transition and is captured in the Land\'{e} $g$-factor.

Determining the Land\'{e} $g$-factor for a transition can be trivial for some species, but difficult for others. Paramagnetic species have magnetic moments due to (an) unpaired electron(s). In atomic species, the electrons are endowed with magnetic moments due to both the orbital angular momentum and the internal spin angular momentum. For light nuclei, and singly unpaired electrons, quantum numbers of the orbital and spin angular momenta are well-defined, leading to the relatively easy evaluation of the total magnetic moment due to the electron. For molecular radicals, things start to get more complicated. Electron angular momenta couple with the molecular rotation (and perhaps any nuclear magnetic momenta), resulting in so-called fine structure. For such species, electronic quantum numbers are not well-defined but approximate. In order to compute the $g$-factor for a certain transition, the exact mixing of the electron angular momentum states has to be derived, which usually is possible in the case that the fine-structure spectrum is experimentally known \citep{larsson2019, larsson2020}. Transition-specific $g$-factors derived in this way are highly reliable. Finally, there are some molecules, such as C$_4$H, that have electronic configuration-states that lie close in energy \citep{oyama2020}. For these molecules, fine structure is determined by a mix of the electronic states, coupling to the rotation, and computing magnetic moments due to unpaired electrons becomes highly non-trivial.

For diamagnetic molecules, the Zeeman effect is due to the magnetic moments of the nuclei.  Their sensitivity to the Zeeman effect is significantly less than for paramagnetic species, so that the splitting is only observable in bright maser lines like those of CH$_3$OH and H$_2$O (\S\ref{sec-masers} \& Table~\ref{zee-tab}), with accurate Land\'{e} $g$-factors \citep[e.g.,][]{lankhaar2018}.  
The values of the Zeeman-splitting factors listed in Table~\ref{zee-tab} are at present the most accurate available.  In cases where only an estimate is given, the derivation of an accurate value is a work in progress (Lankhaar, in prep.).


\section{Regions for Study (and their Zeeman Effect Tracers)}
\label{sec-regions}

\subsection{Molecular Clouds}

\begin{SCfigure}[][ht]
\centering
\includegraphics[align=m,width=0.5\columnwidth]{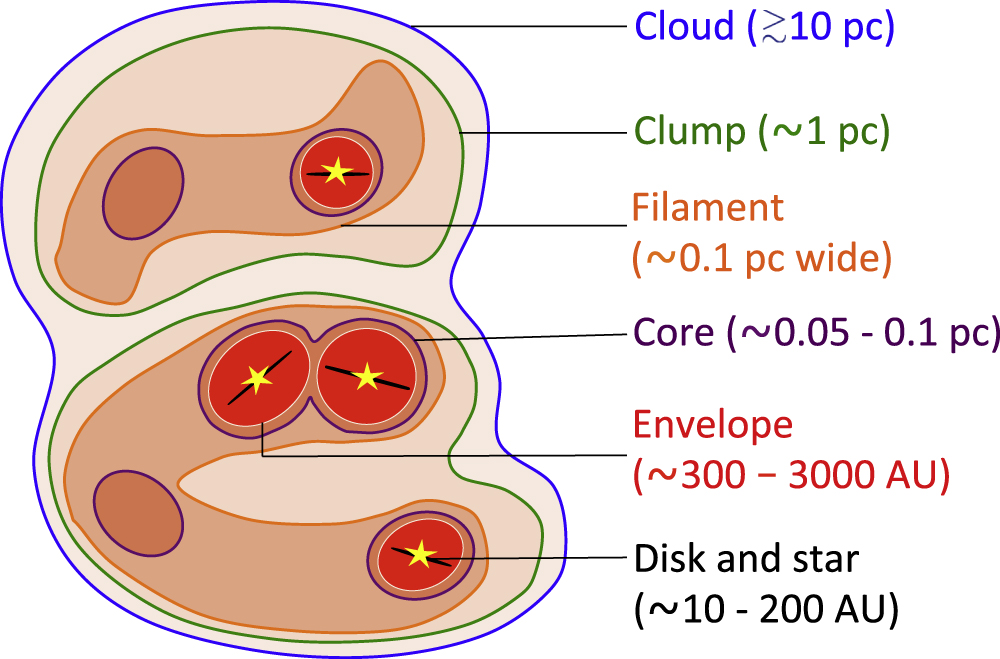}
    \caption{
    A cartoon display of a molecular cloud showing hierarchical structures inside the cloud. The figure shows the cloud, clumps, filaments, cores, envelopes, and protostellar systems that we consider in this study. The image is not drawn to scale \citep[from][]{pokhrel2018}.
    \\
    \\
    \\} 
    \label{fig:cloud}   
\end{SCfigure}

Stars form in dense, compact regions within large molecular clouds ($\geq$ 10 pc).   Molecular clouds are hierarchical, likely due to fragmentation \citep{dobbs2014}.  Within molecular clouds are "clumps" ($\sim$1 pc) which contain elongated "filaments" of width $\sim$0.1 pc, to within a factor of 2 \citep{arzoumanian2011}.   Some filaments converge into "hubs" of size a few $\times$ 0.1 pc, in which clusters of stars may form.  Along the filaments are found "dense cores" of size $\sim$0.05-0.1 pc \citep{difrancesco2007} in which individual protostellar systems (hereafter, protostars) form.  Cores without stars are starless but those that are gravitationally bound are called prestellar cores, although observationally this is difficult to determine.  Starless cores with central densities $> 10^4$ cm$^{-3}$ are likely to be prestellar. Surveys, primarily with Herschel, have revealed that more than 70\% of protostars and prestellar cores are found within filaments \citep{andre2014}.   Protostars are surrounded by "envelopes" of size a few thousand au, from which they directly accrete. Figure~\ref{fig:cloud} presents a not-to-scale cartoon of the hierarchical structure of a molecular cloud as described here.

Dust polarisation maps show that magnetic fields in the gas surrounding dense filaments (sheaths) are typically aligned perpendicular to the main axis of the filament at column densities $N_{\rm H} < 5 \times 10^{21}$ cm$^{-2}$, becoming more aligned at the edges of the filaments as column densities increase above this number \citep{planckxxxv, pattle2023}.  This result leads to the question of whether the fields are shaped by the flow of gas, or whether the fields control the flow of gas, both outside and inside the filaments.  Simulations suggest these alignments primarily occur when the magnetic field is dynamically important, so that the fields control the flow of gas \citep{soler2013}.  Measurements of magnetic field strength would greatly assist in answering this question.  With SKA-Mid, the gas at moderate column density around the dense filaments can be observed in OH emission or absorption depending on the gas temperature \citep{ebisawa2020}. Depending on the exact structure (its smoothness) and temperature of the gas, it may  also be observed in HI, if it is not strongly spatially filtered. 

\subsection{Low-mass Star-forming regions}

Low-mass star formation occurs within cores of size $<10\, \Msun$, both in isolation ($\sim$1-2 cores per clump) and in clustered environments.  While low-mass stars also form in regions of high-mass star formation, here we are focussed on regions without high-mass stars, such as the nearby molecular clouds of the Gould Belt with distances typically $<$ 300 pc (southern sky examples accessible to the SKA include Lupus, Corona Australis, Ophiuchus, \& Chamaeleon).  

The classical picture of relatively isolated star formation starts with a dense starless core threaded by a magnetic field that increases in central density until gravity becomes dominant and the core collapses.  In this simple picture, the evolution of the core can be slow or fast depending on the strength of the magnetic field, but in either scenario the magnetic field evolves into an hour-glass shaped field with the pinch at the location of the protostar (or in the case of a prestellar core, where the protostar is forming, i.e., the density peak).  This field structure has been observed in polarised dust emission in some protostellar cores (Girart 2006) and perhaps in a prestellar core (Kim et al., submitted), but most cores show more complex structures (protostars) or almost parallel fields (prestellar/starless cores). Dense cores are dominated by thermal motions, showing molecular line widths that are usually sub-sonic (and sub-Alf\'venic) but can be trans-sonic in the presence of a protostar.  The line widths are thus narrow and favourable for Zeeman effect observations. 

In low-mass starless/prestellar cores the best tracer of the Zeeman effect is likely to be CCS, as it traces the right densities and is abundant in chemically unevolved regions \citep{suzuki1992}.  Attempts to observe the Zeeman effect using CCS have been made toward TMC-1 and L1498 \citep{guesten1990, koley2022, shinnaga1999}, while surveys find bright CCS toward some cores, with CCS/NH$_3$ ratios suggesting they are chemically young \citep{suzuki1992, foster2009, marka2012, seo2019}.  CH and C$_4$H are promising candidates for Zeeman effect observations toward starless cores, but as indicated earlier, insufficient total power observations of Zeeman sensitive transitions of these molecules exist to know for sure.  

In the envelopes (clumps) around dense cores, OH emission has been used to observe the Zeeman effect, in both starless and protostellar cores \citep{crutcher2000,bourke2001}, but this emission may be resolved out by SKA-Mid, and simulations are required to examine this.  HINSA is a promising probe of magnetic fields in low-mass star-forming regions, as already demonstrated by the FAST results \citep{2022Ching} toward the prestellar core L1544. 
 HINSA has been shown to originate in the cold, well-shielded interiors of dark clouds, and toward low-mass cores, HINSA is often found to have the same non-thermal line width and the same line centroid velocity as the OH emission \citep{2003LG} and generally follows the distribution of the $^{13}$CO and C$^{18}$O emission \citep{2005GL}, all of which trace densities $\sim$10$^{3-4}$ cm$^{-3}$.  
 
\subsection{High-mass Star-forming Regions}

High-mass stars (i.e., ionising, with $M > 8$ \Msun\ and $L \geq 10^3$ \Lsun) form in massive regions of high column density \citep{beuther2007, motte2018, urquhart2024}, the "hubs" discussed earlier where filaments are observed to converge ($\sim$1000 \Msun).   These hubs contain one to a few massive dense cores (MDC) with sizes 0.1-1 pc, from which high-mass protostars form.  High-mass stars always form in a clustered environment with lower mass stars, with a distribution generally following the IMF in slope - but clusters don't always form high-mass stars.  Whether high-mass star formation (HMSF) is analogous to low-mass star formation is an open question \citep{beuther2025}, but there is at least one key difference.  High-mass stars reach the main sequence while still embedded in and accreting from their natal MDC.  Therefore the picture of HMSF is complicated by the competing forces of radiation pressure and accretion.  Observationally it can be difficult to study HMSF as most of these regions lie at larger distances than low-mass regions, they contain protostellar clusters of a few hundred to over 1000 members, making it difficult to disentangle cluster and protostar properties, and for the more evolved regions they contain some form of HII region.   Hubs containing MDCs, before a high-mass protostar forms, have been identified in both cold dust emission at submm/mm wavelengths and as FIR/MIR dark clouds against the Galactic infrared emission (IRDC - Infrared Dark Clouds), with a very strong correlation between the dust emission and absorption, and with column densities (and hence masses) significantly higher than measured in low-mass clumps \& cores.   These regions are usually termed "infrared-dark" or IR-dark MDCs, to distinguish them from the larger IRDC in which they reside.  High-mass protostars that form in MDCs, before they reach the main sequence and produce strong radio continuum emission, are identified via their bright FIR/MIR infrared emission, and so in this evolutionary phase they are called "infrared-bright" (an IR-bright MDC may and usually does reside in an IRDC).  This IR-bright MDC might contain a compact hot molecular core ($>$ 100 K) surrounding the protostar, rich in spectral lines of complex molecules.  As the high-mass protostar grows (accretes) it eventually becomes massive enough to ionise its surroundings, and is now both infrared-bright and detectable in the radio.   It will then evolve and expand from a hyper-compact HII region to an ultra-compact HII region to a compact HII region to the classical HII region which may contain multiple sites of high-mass star formation \citep{churchwell2002, urquhart2024}.

Masers are important tools in the study of high-mass regions, and the presence or absence of certain masers is used as a proxy for their evolutionary status \citep{breen2010, urquhart2024}.
As the high-mass star evolves through the hot molecular core phase into the HII region phases, the maser lines of CH$_3$OH and OH (and H$_2$O) appear under certain physical conditions and as discussed in \S\ref{sec-masers} can provide important information on magnetic field strengths that are complementary to Zeeman effect measurements in non-maser lines.     

All of the tools discussed earlier to observe the Zeeman effect and hence measure magnetic field strengths are applicable to HMSF regions.   In the earliest phase before the onset of ionising radiation, CCS is a promising candidate for Zeeman observations, showing moderate total intensity emission in regions of high-density cold gas \citep{sakai2008, worthen2025}.  Both CH and C$_4$H are promising candidates in this regard, but as with low-mass regions, surveys of total intensity are required before dedicated Zeeman observations are attempted.  SO as both a high density and shock tracer is also promising for HMSF regions, in particular the early cold phase and the hot molecular core phase, and surveys in this line should be undertaken \citep{fontani2025}.   Throughout the HII region phases, absorption line observations of HI and OH are important, as they have already produced many of the Zeeman effect detections to date (Fig~\ref{fig:relations}).  These absorption studies are sensitive to the thin PDR layer between the HII region (densities $\sim10^4$ cm$^{-3}$) and the surrounding molecular cloud and thus provide information on the field at the edge of the HII region.  Within the HII region itself Zeeman effect observations of RRLs may be a tool for SKA (Low and Mid) studies of the ionised gas.

\subsection{Protoplanetary Disks}

Understanding the structure and strength of magnetic fields in protoplanetary disks (PPDs) is important for constraining theories of disk evolution that are strongly based on the favoured mechanism of magnetohydrodynamic (MHD) instabilities \citep{teague2025}.  While dust polarimetry is now routinely used to trace field structure on different size scales in molecular clouds (\S\ref{sec-dust}), in PPDs the continuum polarisation is often due to other physical effects such as self-scattering.  While line-broadening due to the Zeeman effect holds promise for inferring field strengths in some disks (\S\ref{sec-broad}), it would be preferable to measure field strengths more directly via the Zeeman effect.  In non-maser lines, this will be (and has been) challenging for two main reasons.  

First, the unpolarised total intensity of SKA Zeeman sensitive tracers is typically not large in PPDs, for transitions observed at higher frequencies (SO, CCS), and so their sensitivity to the Zeeman effect is low, although many transitions of Zeeman sensitive species have not yet been observed toward PPDs at SKA frequencies to know if that holds true at lower-frequencies.    While RRLs may be important for studies of disks that are externally radiated e.g., near to OB stars such as the Orion proplyds, as with RRLs in other regions their total intensity may be too low to be useful for Zeeman observations of PPDs (Garufi et al. 2026).  HI at 21-cm has not yet been detected in a PPD, and so is unlikely to be a useful Zeeman tool.  

Second, the field geometry in PPDs is complex, with both poloidal and toroidal components, and field reversals, present in even a synthesised beam from an interferometer \citep{mazzei2020}.  Thus it may be that the Zeeman effect is only detectable in PPDs with a particular geometry, such as face on, where the field is aligned with the observer, although initial observations at ALMA frequencies of the face-on disk of TW Hya have proved unsuccessful \citep[e.g.,][]{vlemmings2019}.  


\section{Synergies with Other $B$-field Tracers and Other Facilities}
\label{sec-other}

\subsection{Dust Polarimetry}
\label{sec-dust}
Dust polarimetry is a widely-used technique for mapping plane-of-sky magnetic field direction (\bposdir), using the preferential alignment of interstellar dust grains with respect to the local magnetic field direction \citep{davisgreenstein1951,andersson2015}.  This method can be used both in the optical/near-IR by measuring the polarization of background starlight caused by selective dust extinction, or in the far-IR to mm by observing linearly polarized thermal dust emission.
Current- and recent-generation (sub)mm and far-IR facilities with polarimetric capabilities have over the last decade significantly enhanced our understanding of the morphology and dynamic importance of magnetic fields in star-forming regions \citep{pattle2023}. Starlight polarization measurements combined with dust reddening measurements and GAIA-measured parallax distances, have been used to perform polarization tomography, decomposing the observed polarization into multiple ``screens''  of polarized dust, each with a different distance along the line of sight and magnetic field orientation \citep[e.g.][]{panopoulou2019,Angarita2025,Panopoulou2025}.  In the SKA era new polarized emission surveys from CCAT/PrimeCam \citep{CCAT2023} and Simons Observatory \citep{Ade2019,Clancy2023} will map most of the Southern Sky providing arcminute or better resolution \bposdir\ maps of tens of thousands of dense cold clumps and filaments. 
Proposed (sub)mm/far-IR polarization-sensitive cameras may provide significant improvements in both angular resolution and sensitivity over current-generation instrumentation, such as those proposed for the Probe far-Infrared Mission for Astrophysics (PRIMA)  \citep{burgarella2024}, for the Atacama Large-Aperture Submillimeter Telescope (AtLAST; \citealt{mroczkowski2025}), and future ALMA wideband upgrades \citep{carpenter2023}. 
Therefore, SKA Zeeman observations will arrive in a context where plane-of-sky magnetic field direction will have been mapped with high sensitivity and angular resolution over a large fraction of the sky \citep[e.g.,][]{klaassen2024,pattle2025}.

While dust polarimetry can provide rapid mapping of plane-of-sky magnetic field morphology, it can only indirectly measure magnetic field strengths using the DCF method \citep{davis1951,chandrasekhar1953}.  This method, while widely used, is subject to many known observational biases, and despite significant recent improvement efforts is likely to be accurate only to within a factor of a few \citep[e.g.,][]{liu2022, pattle2023}.  Calibration of the DCF method could be significantly improved through benchmarking against Zeeman measurements in as wide a range of astrophysical environments as possible.  However, previous attempts to do so have been limited by the prohibitively small number of Zeeman measurements presently available, and the need to ensure that the Zeeman and DCF measurements trace the same material \citep{poidevin2013}.  Moreover, this benchmarking must be performed statistically since the Zeeman effect and dust polarimetry trace orthogonal magnetic field components \citep{crutcher2004}.  The SKA will provide an unprecedentedly large sample of Zeeman-measured magnetic field strengths across a wide range of gas densities, offering the possibility of accurately calibrating this widely used technique.

Dust polarization also encodes information on the inclination angle of the magnetic field with respect to the plane of the sky, $\gamma$, through the polarization fraction, which is proportional to $\cos^2\gamma$ \citep{lee_draine1985,fiege2000}. If the dust properties can be assumed to be uniform across the cloud then measurements of the polarization fraction can be used to estimate the average inclination angle of a cloud \citep{chen_cy2019,sullivan2021}. More recently \cite{hoang2024} have developed a modification of the \cite{chen_cy2019} method that also includes variations due to grain shape and alignment efficiency. Combining Zeeman measurements of \blos\ and constraints on the inclination angle due to dust polarization can be used to estimate the 3D magnetic field strength. 

\subsection{Faraday Rotation}
A complementary technique for studying the line-of-sight magnetic field is the Faraday rotation of linearly polarized radiation in a magnetized plasma. In a simple scenario, when linearly polarized light from a background source passes through a Faraday-rotating medium, the amount of rotation ($\Delta \theta$) has a linear relationship with wavelength squared ($\lambda^2$), and the slope is referred to as rotation measure (RM~$= \Delta\theta / \Delta \lambda^2$). \cite{Vanecketal2023} provide a consolidated catalog of RMs to-date with more than 55,000 observations. This number is already being significantly increased with the upcoming rotation measure maps from the Polarisation Sky Survey of the Universe's Magnetism~\citep[POSSUM;][]{Jungetal2024, Vanderwoudeetal2024, gaensler2025} and SPICE-RACS
\citep[with over 250,000 RMs;][]{Thomsonetal2023SPICERACS} with the SKA precursor ASKAP.

Faraday rotation of these background sources can measure the line-of-sight magnetic fields through molecular clouds using the MC-BLOS technique \citep{Tahanietal2018, Tahanietal2025}. The technique has two components for determining line-of-sight magnetic fields: 1) direction determination, and 2) strength determination.  The direction determination incorporates an on-off approach for determining the RM induced by a cloud's magnetic field at each RM point. The strength determination utilizes column density (extinction) maps and a chemical evolution code.  The application of the MC-BLOS technique to some nearby molecular clouds yielded results consistent with Zeeman observations and found a \blos\ reversal across the Orion A, California, and Perseus molecular clouds \citep{Tahanietal2018}. This reversal was previously detected around the Orion A cloud using HI Zeeman observations. The follow-up 3D field reconstruction studies found that an arc-shaped field morphology around the cloud is causing this observed reversal \citep{Tahani2022, Tahanietal2022P, Tahanietal2022O}.  Future RM studies, made with the SKA telescopes, can be strongly complemented by atomic and molecular Zeeman observations (see chapter by \cite{Tahani2026.SKA}). 

\subsection{The Goldreich-Kylafis Effect}

In the presence of a magnetic field, molecular rotational levels may split into magnetic sub-levels.  Radiation emitted by transitions between these sub-levels may be polarized either parallel or perpendicular to the plane-of-sky magnetic field direction in the presence of an anisotropic velocity gradient and/or radiation field.  This spectral line linear polarisation is known as the Goldreich-Kylafis (GK) effect \citep{goldreich1981, heiles1991}, and should be observable in many spectral lines with optical depth close to unity.  The GK effect offers the possibility of a velocity-resolved tracer of plane-of-sky magnetic field morphology, but interpretation is complicated by the uncertainty on polarization direction, and due to the weakness of the effect, it has thus far been observed only in (sub)mm lines  \citep{girart1999, ching2016}, and not at all at cm wavelengths accessible to the SKA.  In regions where both the GK and Zeeman effect are detected they offer the promise of complementary (plane-of-sky and line-of-sight) velocity-resolved information on the magnetic field, but this promise awaits realisation.   

\subsection{Line Broadening due to the Zeeman Effect}
\label{sec-broad}

Although the main signature of the Zeeman effect on spectral lines is a frequency offset of the RCP and LCP components in proportion to the LOS field strengths and Land\'e $g$-factor, the Zeeman effect leaves additional
signatures in spectral lines \citep{lankhaar2023}. First, the shifting of polarization modes also produces linear polarization, which scales with the square of the plane-of-sky field component. The expected linear polarization fraction is of order $\sim \Delta Q[V/I]^2 [B_{\mathrm{POS}}^2/B_{\mathrm{LOS}}^2]$, where $\Delta Q$ is a proportionality factor, that can become very large for lines with transitions that are associated with large angular momenta \citep{lankhaar2023}. Usually, linear polarization fractions are low, but they can become high for sources with a favorable geometry and for transitions that are sensitive to linear polarization. Second, Zeeman splitting broadens the spectral line in proportion to the square of the total field strength: that is of order $\Delta v_Z / \mathrm{FWHM}\sim \bar Q [V/I]^2 [B_{\mathrm{TOT}}^2/B_{\mathrm{LOS}}^2]$, where $\bar Q$ is the broadening factor, which can become very large for lines with transitions that are associated with large angular momenta. Measuring the Zeeman broadening across a manifold of CN $N=1-0$ transitions with varying Zeeman sensitivities has successfully constrained the magnetic field strength of the protoplanetary disk TW Hya \citep{teague2025}. If multiple Zeeman signatures (circular polarization, linear polarization, and line broadening) can be measured simultaneously, they together enable a full 3D characterization of the magnetic field from a single tracer \citep{lankhaar2023}.

\subsection{Zeeman Tracers at non-SKA Frequencies}

Zeeman effect observations at frequencies higher than those covered by the SKA have been successfully performed using the septet of CN lines at 113 GHz with the IRAM 30-m telescope \citep{crutcher1996, crutcher1999c}, and with H$_2$O masers (\S\ref{sec-masers}).  There are a number of transitions in the 13, 7 and 3 mm bands that are potential Zeeman tracers of high-density and shocked gas, notably CCS at 22, 33, and 45 GHz (including Zeeman observations in the 22 GHz and 45 GHz lines; \cite{shinnaga1999, levin2001, nakamura2019, koley2022}), SO at 30, 86, and 99 GHz, and the C$_2$H triplet at 87 GHz \citep{robishaw2008, crutcher2019}. Attempts have been made to detected the Zeeman effect in the CN lines with ALMA, without success. The CCS and SO lines, having Zeeman sensitivity transitions at SKA frequencies, offer the potential of measuring the field strength in multiple transitions of the same molecule.  There now exists a number of ALMA surveys in the 3-mm band covering some of these lines, such as the ALMA-ATOMS survey of high-mass star forming clumps \citep{liu2020} which provide catalogs for identifying Zeeman candidates.  Future surveys with ALMA and eventually AtLAST will identify new candidates, and deep Zeeman observations should be undertaken with both facilities, while keeping in mind the polarisation sensitivity limits of ALMA (see the ALMA Technical Handbook).

\section{Metrics}
\label{sec-metrics}
The review by \cite{pattle2023} discusses in detail how observations of magnetic fields are used to understand their importance in molecular cloud formation and evolution, and in star formation.  
Here we focus on a few key areas where magnetic field strength measurements via the Zeeman effect with the SKA will provide crucial information for advancing our understanding of the role of magnetic fields in star formation.  

As noted earlier, the number of {\it detections} of the Zeeman effect in Fig.~\ref{fig:relations} is small, and the main goal is to increase the number of measurements on multiple size-scales and densities within molecular clouds.   With the SKA this means undertaking Zeeman effect observations on scales of clumps (1 pc and 10$^3$ cm$^{-3}$) to the scales and densities of envelopes around protostellar systems, and perhaps the disks around protostars (Fig.~\ref{fig:cloud}).   While the $B-n$ and $B-N$ data in Fig.~\ref{fig:relations} suggests that molecular clouds on most size scales are magnetically supercritical and undergoing self-similar collapse ($\kappa \sim 0.65$), the small number of data points, and arguments suggesting issues with both the data (density estimates) and analysis, means these results are not conclusive, and could even be consistent with a $B-n$ index of $\kappa\sim$0.5 \citep{tritsis2015}.  The relationships may even suggest fields are about critical for  more realistic geometries such as non-uniform densities \citep{myers2021}. It is not clear if these relationships hold within individual regions, as maps and measurements at different densities are mostly lacking.  Attempts to infer the $B-n$ relationship within individual clouds using the DCF method suggests the relationship could exhibit a shallower index of 0.5 \citep[e.g.,][]{2015Li}, while global DCF results are broadly consistent with $\kappa\sim$0.6, with significant scatter \citep{pattle2023}.  

Another measure of the importance of magnetic fields is the mass-to-magnetic flux ratio ($M/\Phi \propto N_{\rm H}/B$), which is a measure of the ratio of gravitational to magnetic pressure \citep{mckee1993}, and provides a straightforward way to determine whether magnetic fields are strong enough to support clouds against gravity.  The critical mass-to-flux ratio is the ratio for which the gravitational and magentic energies are in equilibrium, $(M/\Phi)_{\rm crit}$ = ${\rm c}_\Phi/\sqrt{G}$, where the numerical value ${\rm c}_\Phi$ depends on the cloud geometry, being $\approx0.12$ for a uniform spherical cloud with flux-freezing, and $\approx0.16$ for an isothermal sheet with a constant mass-to-flux ratio \citep{bourke2001}.  The mass-to-flux ratio is usually expressed in terms of the critical value, 
\begin{equation}
    \mu_{\Phi} = \frac{(M/\Phi)}{(M/\Phi)_{\rm crit}} = \frac{\sqrt{G}}{{\rm c}_\Phi} \frac{N_{\rm H}}{B}
\end{equation}
with correction factors of a few for geometry \citep{bourke2001}.  Values of $\mu_{\Phi} > 1$ indicate the field cannot prevent gravitational collapse, and the region is called magnetically supercritical, while values $<$ 1 indicate a magnetically subcritical region (Fig~\ref{fig:relations}(b)).  Experimentally values of $\mu_{\Phi}\, \la\, 2$ are generally considered to be consistent with being approximately critical, and thus magnetic fields play a role in regulating cloud collapse.  While many regions observed to date are formally supercritical by a factor of 2-3, they may still slow collapse without preventing it \citep{myers2021}.   

Because Zeeman observations use spectral lines, information on line width and hence velocity dispersion is obtained, while the field strength measurements allow for the determination of the Alfv\'en velocity.  This information enables the determination of the Alfv\'en Mach number $M_A$, the ratio of non-thermal velocity dispersion and Alfv\'en velocity, which is a measure of the relative importance of kinetic and magnetic energies.  In general, values of $M_A < 1$ (sub-Alfv\'enic) implies that magnetic fields direct gas motions, while $M_A > 1$ indicates the gas guides the fields (super-Alfv\'enic).  The relative importance of magnetic fields in the support and evolution of molecular clouds on different scales can be judged by determining how close the different energy densities (magnetic \md, gravitational \gd, and kinetic \kd) are to equipartition \citep{myers1988, mckee1993}.  By comparing maps of magnetic field strength distribution with velocity structure maps and density distribution maps (column and number densities), one can examine the relative contributions of \md, \gd, and \kd\ from diffuse envelopes to dense cores. Such comparisons will help quantify how energy is transferred and dissipated during star formation.


\section{Predictions for measuring magnetic field strengths with the SKA}

The 3-$\sigma$ sensitivity of Zeeman effect observations to a particular value of the magnetic field strength can be estimated using the following equation \citep{troland1982}:

\begin{equation}
B_{min} [\mu {\rm G}]= 2\ \left[\frac{1}{Z}\right] \left[\frac{FWHM}{1\ {\rm Hz}}\right]  \left[\frac{\Delta I}{I}\right] 
\end{equation}

where $I$ is the line intensity, $\Delta I$ is the rms (or equivalently, $\Delta I/I$ is the inverse of the S/N ratio), the line width is in Hz, and the spectral resolution allows for 6 channels across the line FWHM.  

The combination of sensitivity and resolution offered by the SKA covering a number of Zeeman-sensitive transitions (Table~\ref{zee-tab}) promises to significantly increase in the number of Zeeman effect detections in the different types of star-forming environments discussed in \S\ref{sec-regions} over a range of densities and size scales.  Below we present some case studies using the SKA-Mid AA4 to illustrate this.  

\subsection{Case study 1: OH Zeeman effect in Orion B - the Gold standard}
\begin{figure}[t!]
    \centering
    \includegraphics[width=0.9\columnwidth]{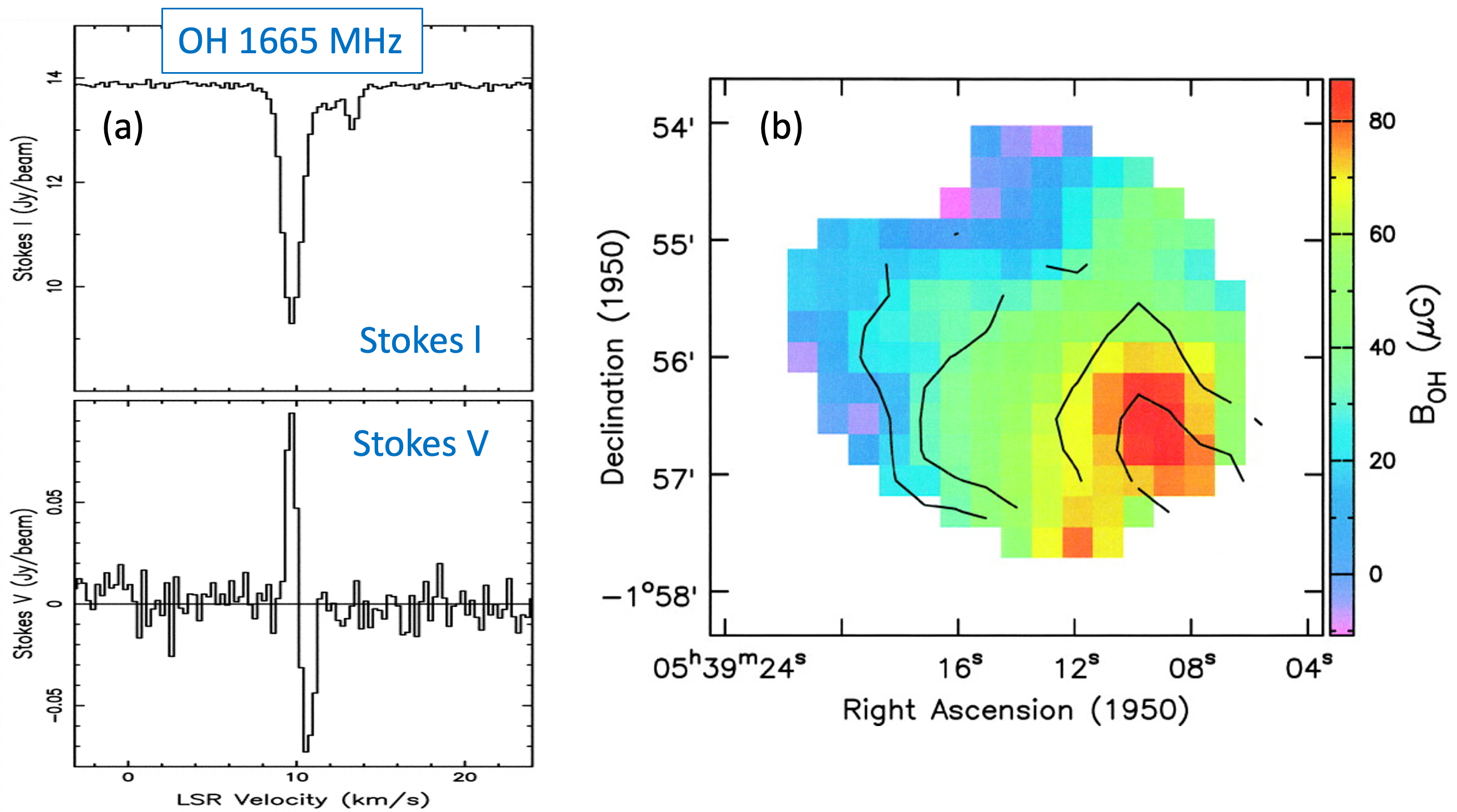}
    \caption{OH Zeeman measurements toward NGC 2024 using the VLA with a beam size of $\sim$60\arcsec\; Figure adapted from \cite{crutcher1999a}. Panel (a) shows the OH 1665 Stokes I and V spectra at the position of the peak field, while (b) shows the map of derived field strength.  SKA-Mid will obtain similar results in a fraction of the time with a $\sim$6\arcsec\ beam.} 
    \label{fig:orionB}
\end{figure}

The Zeeman effect has been well-detected in OH absorption toward the HII region Orion B (NGC 2024; 380 pc) with a number of facilities \citep[Nan\c cay, Parkes, Green Bank 140-ft, VLA;][]{ck1983, bourke2001, crutcher1999a}. 
Using the VLA, \cite{crutcher1999a} were able to map the Zeeman effect in Orion B down to $\sim$10 $\mu$G, with a peak value of around 90 $\mu$G, with $\sim$60\arcsec\ resolution (Fig.~\ref{fig:orionB}).  Their observations, combining the VLA C+D configurations, had an on-source integration time of $\sim$21 hrs, resulting in a per-channel rms noise of $\sim$6 mJy/beam (0.15 km/s channels).  With MeerKAT this noise can be achieved in $\sim$0.5 hr with the same beam size (but with 0.29 km/s channels), and because of its array layout, that same sensitivity is still achieved with only a 10\arcsec\ beam.   Ultimately, with the SKA-Mid a similar sensitivity can be obtained in a $\sim$6\arcsec\ beam (0.012 pc) in only 5 minutes (a factor $\sim10$ improvement in resolution).  With SKA-Mid, consideration may need to be given as to whether such a short integration provides suitable (u,v) coverage for an extended source, and the time needed for polarisation calibrations. This analysis is the subject of a separate effort beyond this paper. Deeper SKA-Mid observations will allow for mapping of the field down to $\mu$G levels on these scales.   

\subsection{Case study 2: OH Zeeman effect in DR21}

\begin{figure}[t!]
    \centering
    \includegraphics[width=0.9\columnwidth]{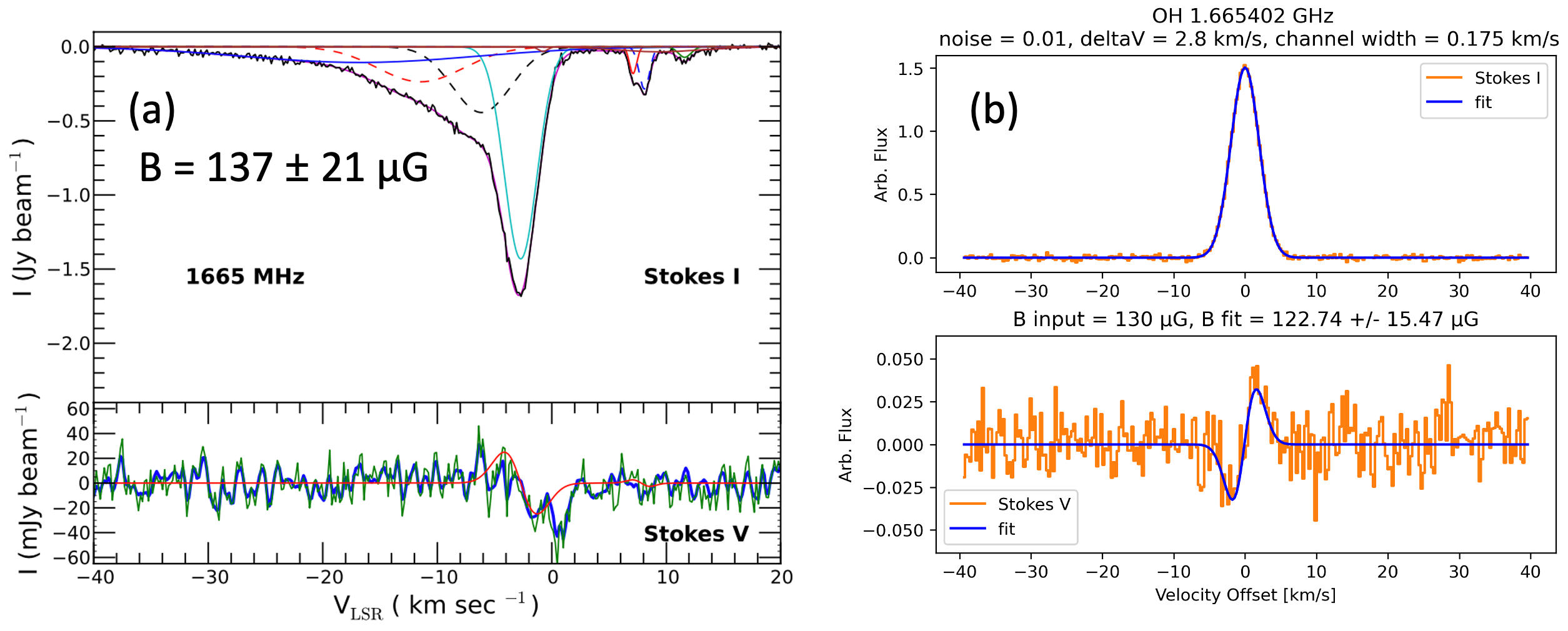}
    \caption{(a) OH Zeeman measurements toward DR21 using the VLA \citep[adapted from][]{koley2021}. (b) Simulated spectra of the DR21 result, confirming the robustness of the VLA result (the simulation is agnostic to whether the input profile is "emission" or "absorption" as the ultimate result is the same).} 
    \label{fig:dr21}
\end{figure}

Most Zeeman detections in thermal (i.e., non-maser) lines are not as clear as the Orion B case.  An excellent example is the recent detection of the OH Zeeman effect toward DR21 \citep{koley2021}, one of the few new detections since the Crutcher review \citep{crutcher2012}.  \cite{koley2021} used the VLA with spectral resolution of 0.175 \kms\ in the D-configuration (beam $\sim30\arcsec$) with an on-source integration time of $\sim$8 hr, resulting in an rms noise of $\sim$10 mJy.  As they observed both OH 18 cm mainlines, they were able to confirm their detection of a field strength of $\sim$130 $\mu$G.  The line profile is complex, requiring at least 5 Gaussian components to fit the main absorption line (negative velocities), with the deepest fitted absorption line giving rise to the Zeeman effect.  The Zeeman signal and fit might not look convincing (Fig.~\ref{fig:dr21}) but the formal results are secure, as confirmed by our simulation with similar parameters.  A similar result could be achieved with the SKA-Mid with only a 6 minute integration (with an rms noise $<$ 5 mJy in a smaller beam $<$ 20\arcsec.)  With significantly more integration time SKA-Mid will be sensitive to much lower field strengths (e.g., 20 $\mu$G in 24 hrs for spectra 10 $\times$ fainter), allowing for mapping of the field toward nearby ($\leq$ 2 kpc) high-mass star-forming regions similar to DR21. 

\subsection{Case study 3: CCS Zeeman effect at 11 GHz in star-forming cores with SKA-Mid}
\label{sec:case3} 

Surveys of the higher-frequency CCS lines (> 11 GHz) suggest that CCS traces mean densities of $3 \times 10^4$ cm$^{-3}$, and shows typical linewidths of $\sim$0.7 \kms\ and $\sim$0.2 \kms\ for low-mass protostellar and prestellar cores, respectively (and $\sim$1.5 \kms\ in infrared dark clouds).  Using this mean density, assuming the source completely fills the beam (reasonable for the relatively shallow inner density profiles of low-mass cores, at least) and shifting the Crutcher relationship ($B \propto n^{0.65}$) down by 0.5 dex (to ensure we are sensitive to a field strength well below the 0.65 line -- see Fig~\ref{fig:relations}), we simulate the S/N and hence time required with SKA-Mid to detect a field of 300 \uG\ and 100 \uG\ (3$\sigma$) for protostellar and prestellar low-mass cores, respectively, with a peak line flux of 0.5~Jy in a 30\arcsec\ beam (equivalent to a $\sim$1~K line observed with the GBT).  In the protostellar case an integration time of $\sim$3 hrs is needed, and $\sim$8 hrs for the prestellar case.  If the line is only 0.2~Jy then the times increase to $\sim$20 hrs and $\sim$65 hrs respectively.  The limited observations of the 22 GHz line using both the GBT with $\sim$30\arcsec\ beam and the VLA with $\sim$4\arcsec\ beam suggest that the VLA filters out most of the single-dish emission, supporting the assumption that the density distribution is relatively flat. The dense inner core of SKA-Mid will alleviate some of the interferometric filtering issues, but quantifying this will require simulated observations with the SKA-Mid layout of star-forming cores whose density profiles are well modelled, which is beyond the scope of this chapter but is a work in progress.  


\subsection{Zeeman Effect Observations with the SKA -- Plans and Preparations}

It is difficult to make clear predictions on the number of new Zeeman detections that will be made with the SKA (AA4) before the surveys below are undertaken.  The expectation is that  several hundred new detections will be made, with maps in the different density tracers in individual regions.  Observations of the Zeeman effect with the SKA will primarily be undertaken with SKA-Mid (Table 1). SKA-Mid's frequency coverage and instantaneous bandwidth will allow for multiple species to be observed in a single observation, e.g. the 13 GHz lines of OH and SO, the 9 GHz lines of C$_4$H and methanol, and the 11 GHz line of CCS in Band 5b. Before SKA-Mid is operating in steady state (AA4) a number of preparatory observations are needed, some of which have started, while others are not yet planned.   With MeerKAT, observing programs are underway to observe the Zeeman effect in OH, starting with Orion B and including regions with bright OH lines (typically in absorption against HII regions). The survey of the Galactic plane in OH with ASKAP \citep[GASKAP-OH;][]{gaskapoh2024} is about to start its main program, which will provide many new targets for OH Zeeman observations.

FAST has made progress in Zeeman observations using HINSA \citep{2022Ching}, but its northern sky coverage and lack of significant overlap with the sky visible to the SKA limits its utility for synergistic observations.   GASKAP-HI \citep{gaskaphi2013} will provide both a sample of traditional HI absorption (against HII regions) and HINSA candidates, for SKA-Mid Zeeman follow-up.  As noted earlier, simulations are required to determine the extent to which HINSA is possible with the SKA-Mid.  Surveys that include RRLs are also needed, and these may have to wait for proposed complementary surveys proposed for the SKA telescopes (see chapters by \cite{Karska2026.SKA}, \cite{Salas2026.SKA}, \& \cite{Traficante2026.SKA}).

Our knowledge of the distribution and brightness of the other thermal tracers discussed earlier (CH, CCS, C$_4$H, SO) is extremely limited, particularly at SKA frequencies, and so surveys of these lines are needed before dedicating serious SKA time to Zeeman observations.  As these are expected to be high-density tracers, targeted survey observations can be undertaken, for example of dense cores in star-forming regions in the Gould Belt and the Galactic Plane, as revealed by large-area (sub)mm/far-IR line and dust-continuum surveys with e.g., Planck, Herschel, APEX, Mopra, and the JCMT.  These surveys can be undertaken with SKA-Mid AA* for CCS and C$_4$H, as no comparable facility will exist prior with sufficient sensitivity and frequency coverage for this undertaking.   For CH at 700 MHz a targeted survey with ASKAP could be undertaken once the bulk of its current large programs are completed, or with SKA-Mid AA*.  

Beyond the current SKA baseline (AA4), an increase in the frequency coverage to SKA-Mid above 15 GHz would give access to the CCS and H$_2$O lines at 22 GHz, and the C$_4$H lines at 19 GHz, while any increase in sensitivity will allow for new Zeeman effect detections.  Being able to observe the Zeeman effect in both the 11 GHz and 22 GHz CCS lines would provide confirmation of a detection in either line. An increase in both angular resolution (VLBI baselines) and sensitivity for SKA-Mid would be advantageous for Zeeman effect observations with masers, enabling statistical studies of magnetic field strength throughout a significant volume of the Galaxy. 

In summary, the combination of sensitivity, resolution, and frequency coverage offered by the SKA will enable the utilization of many Zeeman effect tracers for magnetic field studies in ways not possible with existing facilities. The ability to map regions both near and far over a wide range of densities and spatial scales will greatly advance our understanding the role of magnetic fields in star formation and molecular cloud evolution.  Combining maps of individual sources and statistical results from large ensembles, with complementary observations on larger scales and lower densities with large single-dish telescopes, and using other techniques (\S\ref{sec-other}), will be crucial to making progress in this difficult area.   One key advantage of the SKA is its ability to match beams at different frequencies, as Zeeman observations represent an averaged result over both the telescope’s angular resolution and the regions sampled by the tracer. For instance, even at the same angular resolution and position, Zeeman measurements using HI, OH, and CCS would trace magnetic fields in progressively denser regions from diffuse envelopes to dense cores along the line of sight.  To disentangle these effects, chemical and radiative transfer modeling of multi-tracer Zeeman data from SKA and other telescopes will be essential to reconstruct the three-dimensional magnetic field within a source.


\clearpage 
\bibliographystyle{abbrvnat-maxbibnames4}

{\large\bf Acknowledgements}

AMJ acknowledges the support of the Max Planck Society and SFB~1601.
KP is a Royal Society University Research Fellow, supported by grant number URF\textbackslash R1\textbackslash 211322.

\bibliography{zeeman} 

\end{document}

%% file: zeeman.bib
@incollection{Tahani2026.SKA, author = {Mehrnoosh Tahani and author2 and author3 and author4 and author5},title = {},year = {2026},publisher = {},note = {arXiv search: Report number AASKAII/Tahani01},booktitle = {Advancing Astrophysics with the SKA -- II (AASKAII)}}

@incollection{Karska2026.SKA, author = {Agata Karska and author2 and author3 and author4 and author5},title = {},year = {2026},publisher = {},note = {arXiv search: Report number AASKAII/Karska01},booktitle = {Advancing Astrophysics with the SKA -- II (AASKAII)}}

@incollection{Salas2026.SKA, author = {Pedro Salas and author2 and author3 and author4 and author5},title = {},year = {2026},publisher = {},note = {arXiv search: Report number AASKAII/Salas01},booktitle = {Advancing Astrophysics with the SKA -- II (AASKAII)}}

@incollection{Traficante2026.SKA, author = {Alessio Traficante and author2 and author3 and author4 and author5},title = {},year = {2026},publisher = {},note = {arXiv search: Report number AASKAII/Traficante01},booktitle = {Advancing Astrophysics with the SKA -- II (AASKAII)}}

@incollection{Robishaw2026.SKA, author = {Timothy Robishaw and author2 and author3 and author4 and author5},title = {},year = {2026},publisher = {},note = {arXiv search: Report number AASKAII/Robishaw01},booktitle = {Advancing Astrophysics with the SKA -- II (AASKAII)}}

@incollection{Rygl2026.SKA, author = {Kazi L. J. Rygl and author2 and author3 and author4 and author5},title = {},year = {2026},publisher = {},note = {arXiv search: Report number AASKAII/Rygl01},booktitle = {Advancing Astrophysics with the SKA -- II (AASKAII)}}

@ARTICLE{alves2012,
       author = {{Alves}, F.~O. and {Vlemmings}, W.~H.~T. and {Girart}, J.~M. and {Torrelles}, J.~M.},
        title = "{The magnetic field of IRAS 16293-2422 as traced by shock-induced H$_{2}$O masers}",
      journal = {\aap},
         year = 2012,
        month = jun,
       volume = {542},
          eid = {A14},
        pages = {A14},
          doi = {10.1051/0004-6361/201118710},
archivePrefix = {arXiv},
       eprint = {1204.0004},
 primaryClass = {astro-ph.GA},
       adsurl = {https://ui.adsabs.harvard.edu/abs/2012A&A...542A..14A}
}

@ARTICLE{andersson2015,
       author = {{Andersson}, B.-G. and {Lazarian}, A. and {Vaillancourt}, John E.},
        title = "{Interstellar Dust Grain Alignment}",
      journal = {\araa},
         year = 2015,
        month = aug,
       volume = {53},
        pages = {501-539},
          doi = {10.1146/annurev-astro-082214-122414},
       adsurl = {https://ui.adsabs.harvard.edu/abs/2015ARA&A..53..501A}
}

@ARTICLE{lee_draine1985,
       author = {{Lee}, H.~M. and {Draine}, B.~T.},
        title = "{Infrared extinction and polarization due to partially aligned spheroidal grains : models for the dust toward the BN object.}",
      journal = {\apj},
         year = 1985,
        month = mar,
       volume = {290},
        pages = {211-228},
          doi = {10.1086/162974},
       adsurl = {https://ui.adsabs.harvard.edu/abs/1985ApJ...290..211L}
}

@ARTICLE{sullivan2021,
       author = {{Sullivan}, Callista L. and {Fissel}, L.~M. and {King}, P.~K. and {Chen}, C.-Y. and {Li}, Z.-Y. and {Soler}, J.~D.},
        title = "{Characterizing the magnetic fields of nearby molecular clouds using submillimeter polarization observations}",
      journal = {\mnras},
         year = 2021,
        month = may,
       volume = {503},
       number = {4},
        pages = {5006-5024},
          doi = {10.1093/mnras/stab596},
archivePrefix = {arXiv},
       eprint = {2104.04673},
 primaryClass = {astro-ph.GA},
       adsurl = {https://ui.adsabs.harvard.edu/abs/2021MNRAS.503.5006S}
}

@ARTICLE{fiege2000,
       author = {{Fiege}, Jason D. and {Pudritz}, Ralph E.},
        title = "{Polarized Submillimeter Emission from Filamentary Molecular Clouds}",
      journal = {\apj},
         year = 2000,
        month = dec,
       volume = {544},
       number = {2},
        pages = {830-837},
          doi = {10.1086/317228},
archivePrefix = {arXiv},
       eprint = {astro-ph/0005363},
 primaryClass = {astro-ph},
       adsurl = {https://ui.adsabs.harvard.edu/abs/2000ApJ...544..830F}
}

@ARTICLE{hoang2024,
       author = {{Hoang}, Thiem and {Truong}, Bao},
        title = "{Probing 3D Magnetic Fields Using Thermal Dust Polarization and Grain Alignment Theory}",
      journal = {\apj},
         year = 2024,
        month = apr,
       volume = {965},
       number = {2},
          eid = {183},
        pages = {183},
          doi = {10.3847/1538-4357/ad2a56},
archivePrefix = {arXiv},
       eprint = {2310.17048},
 primaryClass = {astro-ph.GA},
       adsurl = {https://ui.adsabs.harvard.edu/abs/2024ApJ...965..183H}
}

@ARTICLE{chen_cy2019,
       author = {{Chen}, Che-Yu and {King}, Patrick K. and {Li}, Zhi-Yun and {Fissel}, Laura M. and {Mazzei}, Renato R.},
        title = "{A new method to trace three-dimensional magnetic field structure within molecular clouds using dust polarization}",
      journal = {\mnras},
         year = 2019,
        month = may,
       volume = {485},
       number = {3},
        pages = {3499-3513},
          doi = {10.1093/mnras/stz618},
archivePrefix = {arXiv},
       eprint = {1810.10020},
 primaryClass = {astro-ph.GA},
       adsurl = {https://ui.adsabs.harvard.edu/abs/2019MNRAS.485.3499C}
}

@ARTICLE{panopoulou2019,
       author = {{Panopoulou}, Georgia V. and {Tassis}, Konstantinos and {Skalidis}, Raphael and {Blinov}, Dmitriy and {Liodakis}, Ioannis and {Pavlidou}, Vasiliki and {Potter}, Stephen B. and {Ramaprakash}, Anamparambu N. and {Readhead}, Anthony C.~S. and {Wehus}, Ingunn K.},
        title = "{Demonstration of Magnetic Field Tomography with Starlight Polarization toward a Diffuse Sightline of the ISM}",
      journal = {\apj},
         year = 2019,
        month = feb,
       volume = {872},
       number = {1},
          eid = {56},
        pages = {56},
          doi = {10.3847/1538-4357/aafdb2},
archivePrefix = {arXiv},
       eprint = {1809.09804},
}

@ARTICLE{smits2025,
       author = {{Smits}, Derck P. and {Fallon}, Paul},
        title = "{First Detection of Circular Polarization in 4.7 GHz Excited OH Masers}",
      journal = {\apj},
         year = 2025,
        month = sep,
       volume = {990},
       number = {2},
          eid = {193},
        pages = {193},
          doi = {10.3847/1538-4357/adf639},
archivePrefix = {arXiv},
       eprint = {2508.05450},
 primaryClass = {astro-ph.GA},
       adsurl = {https://ui.adsabs.harvard.edu/abs/2025ApJ...990..193S}
}

@ARTICLE{gaensler2025,
       author = {{Gaensler}, B.~M. and {Heald}, G.~H. and {McClure-Griffiths}, N.~M. and {Anderson}, C.~S. and {Van Eck}, C.~L. and {West}, J.~L. and {Thomson}, A.~J.~M. and {Leahy}, J.~P. and {Rudnick}, L. and {Ma}, Y.~K. and {Akahori}, Takuya and {G{\"u}rkan}, G. and {Landecker}, T.~L. and {Mao}, S.~A. and {O'Sullivan}, S.~P. and {Raja}, W. and {Sun}, X. and {Vernstrom}, T. and {Baidoo}, Lerato and {Carretti}, Ettore and {Taylor}, A.~R. and {Willis}, A.~G. and {Osinga}, Erik and {Livingston}, J.~D. and {Alexander}, E.~L. and {Alonso-L{\'o}pez}, David and {Amaral}, A.~D. and {An}, T. and {Bracco}, Andrea and {Bradbury}, S. and {Br{\"u}ggen}, Marcus and {Eswaraiah}, Chakali and {En{\ss}lin}, Torsten and {Galvin}, T.~J. and {Haverkorn}, Marijke and {Hopkins}, A.~M. and {Hutschenreuter}, Sebastian and {Ideguchi}, Shinsuke and {Jaswanth}, S. and {Jung}, S. Lyla and {Kaczmarek}, J.~F. and {Kothes}, Roland and {Lazarevi{\'c}}, Sanja and {Leahy}, Denis and {Loi}, Francesca and {Marvil}, Joshua R. and {Norris}, Ray and {Pandhi}, Ayush and {Price}, Jason M. and {Riseley}, C.~J. and {Ryder}, P. and {Seta}, Amit and {Shaw}, Vasundhara and {Shen}, A.~X. and {Sobey}, C. and {Stil}, J. and {Stuardi}, Chiara and {Upasana}, Gupta and {Vanderwoude}, Shannon and {Velovi{\'c}}, Velibor},
        title = "{The Polarisation Sky Survey of the Universe's Magnetism (POSSUM): Science goals and survey description}",
      journal = {\pasa},
         year = 2025,
        month = jun,
       volume = {42},
          eid = {e091},
        pages = {e091},
          doi = {10.1017/pasa.2025.10031},
archivePrefix = {arXiv},
       eprint = {2505.08272},
 primaryClass = {astro-ph.GA},
       adsurl = {https://ui.adsabs.harvard.edu/abs/2025PASA...42...91G}
}

@ARTICLE{green2012,
       author = {{Green}, J.~A. and {McClure-Griffiths}, N.~M. and {Caswell}, J.~L. and {Robishaw}, T. and {Harvey-Smith}, L.},
        title = "{MAGMO: coherent magnetic fields in the star-forming regions of the Carina-Sagittarius spiral arm tangent}",
      journal = {\mnras},
         year = 2012,
        month = oct,
       volume = {425},
       number = {4},
        pages = {2530-2547},
          doi = {10.1111/j.1365-2966.2012.21722.x},
archivePrefix = {arXiv},
       eprint = {1207.3550},
 primaryClass = {astro-ph.GA},
       adsurl = {https://ui.adsabs.harvard.edu/abs/2012MNRAS.425.2530G}
}

@ARTICLE{green2015,
       author = {{Green}, J.~A. and {Caswell}, J.~L. and {McClure-Griffiths}, N.~M.},
        title = "{Excited-state hydroxyl maser polarimetry: who ate all the {\ensuremath{\pi}}s?}",
      journal = {\mnras},
         year = 2015,
        month = jul,
       volume = {451},
       number = {1},
        pages = {74-92},
          doi = {10.1093/mnras/stv936},
archivePrefix = {arXiv},
       eprint = {1504.07062},
 primaryClass = {astro-ph.GA},
       adsurl = {https://ui.adsabs.harvard.edu/abs/2015MNRAS.451...74G}
}

@ARTICLE{baudry1998,
       author = {{Baudry}, A. and {Diamond}, P.~J.},
        title = "{VLBA polarization observations of the J=7/2, 13.44 GHz OH maser in W3(OH)}",
      journal = {\aap},
         year = 1998,
        month = mar,
       volume = {331},
        pages = {697-708},
       adsurl = {https://ui.adsabs.harvard.edu/abs/1998A&A...331..697B}
}

@ARTICLE{benson1984,
       author = {{Benson}, J.~M. and {Mutel}, R.~L. and {Gaume}, R.~A.},
        title = "{High-angular-resolution observations of the OH masers in W 51 (Main).}",
      journal = {\aj},
         year = 1984,
        month = sep,
       volume = {89},
        pages = {1391-1397},
          doi = {10.1086/113640},
       adsurl = {https://ui.adsabs.harvard.edu/abs/1984AJ.....89.1391B}
}

@ARTICLE{bourke2001,
       author = {{Bourke}, Tyler L. and {Myers}, Philip C. and {Robinson}, Garry and {Hyland}, A.~R.},
        title = "{New OH Zeeman Measurements of Magnetic Field Strengths in Molecular Clouds}",
      journal = {\apj},
         year = 2001,
        month = jun,
       volume = {554},
       number = {2},
        pages = {916-932},
          doi = {10.1086/321405},
archivePrefix = {arXiv},
       eprint = {astro-ph/0102469},
 primaryClass = {astro-ph},
       adsurl = {https://ui.adsabs.harvard.edu/abs/2001ApJ...554..916B}
}

@ARTICLE{balser2016,
       author = {{Balser}, Dana S. and {Roshi}, D. Anish and {Jeyakumar}, S. and {Bania}, T.~M. and {Montet}, Benjamin T. and {Shitanishi}, J.~A.},
        title = "{Magnetic Field Strengths in Photodissociation Regions}",
      journal = {\apj},
         year = 2016,
        month = jan,
       volume = {816},
       number = {1},
          eid = {22},
        pages = {22},
          doi = {10.3847/0004-637X/816/1/22},
archivePrefix = {arXiv},
       eprint = {1511.07383},
 primaryClass = {astro-ph.GA},
       adsurl = {https://ui.adsabs.harvard.edu/abs/2016ApJ...816...22B}
}

@INPROCEEDINGS{burgarella2024,
       author = {{Burgarella}, Denis and {Ciesla}, Laure and {Sauvage}, Marc and {Dowell}, C. Darren and {Donnellan}, J.~M.~S. and {Foote}, Marc and {Glenn}, Jason and {Meixner}, Margaret and {Baselmans}, Jochem and {Bisigello}, Laura and {Bolatto}, Albert and {B{\'e}thermin}, Matthieu and {Bradford}, Charles M. and {Gruppioni}, Carlotta and {Jellema}, Willem and {Moullet}, Arielle and {Oliver}, Seb and {Pamplona}, Tony and {Pope}, Alex},
        title = "{PRIMA: science cases and requirements for the photometric and polarimetric PRIMAger far-infrared camera}",
    booktitle = {Space Telescopes and Instrumentation 2024: Optical, Infrared, and Millimeter Wave},
         year = 2024,
       editor = {{Coyle}, Laura E. and {Matsuura}, Shuji and {Perrin}, Marshall D.},
       series = {Society of Photo-Optical Instrumentation Engineers (SPIE) Conference Series},
       volume = {13092},
        month = aug,
          eid = {130923B},
        pages = {130923B},
          doi = {10.1117/12.3018024},
       adsurl = {https://ui.adsabs.harvard.edu/abs/2024SPIE13092E..3BB}
}

@ARTICLE{bhatnagar2001,
       author = {{Bhatnagar}, S. and {Nityananda}, R.},
        title = "{Solving for closure errors due to polarization leakage in radio interferometry of unpolarized sources}",
      journal = {\aap},
         year = 2001,
        month = aug,
       volume = {375},
        pages = {344-350},
          doi = {10.1051/0004-6361:20010799},
archivePrefix = {arXiv},
       eprint = {astro-ph/0106348},
 primaryClass = {astro-ph},
       adsurl = {https://ui.adsabs.harvard.edu/abs/2001A&A...375..344B}
}

@INPROCEEDINGS{carpenter2023,
       author = {{Carpenter}, John and {Brogan}, Crystal and {Iono}, Daisuke and {Mroczkowski}, Tony},
        title = "{The ALMA Wideband Sensitivity Upgrade}",
    booktitle = {Physics and Chemistry of Star Formation: The Dynamical ISM Across Time and Spatial Scales},
         year = 2023,
       editor = {{Ossenkopf-Okada}, V. and {Schaaf}, R. and {Breloy}, I. and {Stutzki}, J.},
        month = feb,
        pages = {304},
          doi = {10.48550/arXiv.2211.00195},
archivePrefix = {arXiv},
       eprint = {2211.00195},
 primaryClass = {astro-ph.IM},
       adsurl = {https://ui.adsabs.harvard.edu/abs/2023pcsf.conf..304C}
}

@INPROCEEDINGS{cotton1999,
       author = {{Cotton}, W.~D.},
        title = "{Polarization in Interferometry}",
    booktitle = {Synthesis Imaging in Radio Astronomy II},
         year = 1999,
       editor = {{Taylor}, G.~B. and {Carilli}, C.~L. and {Perley}, R.~A.},
       series = {Astronomical Society of the Pacific Conference Series},
       volume = {180},
        month = jan,
        pages = {111},
       adsurl = {https://ui.adsabs.harvard.edu/abs/1999ASPC..180..111C}
}

@BOOK{thompson2017,
       author = {{Thompson}, A. Richard and {Moran}, James M. and {Swenson}, Jr., George W.},
        title = "{Interferometry and Synthesis in Radio Astronomy, 3rd Edition}",
         year = 2017,
    publisher = {Springer},
          doi = {10.1007/978-3-319-44431-4},
       adsurl = {https://ui.adsabs.harvard.edu/abs/2017isra.book.....T}
}

@ARTICLE{bel1989,
       author = {{Bel}, N. and {Leroy}, B.},
        title = "{Zeeman splitting in interstellar molecules.}",
      journal = {\aap},
         year = 1989,
        month = oct,
       volume = {224},
        pages = {206-208},
       adsurl = {https://ui.adsabs.harvard.edu/abs/1989A&A...224..206B}
}

@ARTICLE{brogan2001,
       author = {{Brogan}, C.~L. and {Troland}, T.~H.},
        title = "{VLA H I and OH Zeeman Observations toward M17}",
      journal = {\apj},
         year = 2001,
        month = oct,
       volume = {560},
       number = {2},
        pages = {821-840},
          doi = {10.1086/322444},
       adsurl = {https://ui.adsabs.harvard.edu/abs/2001ApJ...560..821B}
}

@ARTICLE{cazzoli2017,
       author = {{Cazzoli}, Gabriele and {Lattanzi}, Valerio and {Coriani}, Sonia and {Gauss}, J{\"u}rgen and {Codella}, Claudio and {Ramos}, Andr{\'e}s Asensio and {Cernicharo}, Jos{\'e} and {Puzzarini}, Cristina},
        title = "{Zeeman effect in sulfur monoxide. A tool to probe magnetic fields in star forming regions}",
      journal = {\aap},
         year = 2017,
        month = sep,
       volume = {605},
          eid = {A20},
        pages = {A20},
          doi = {10.1051/0004-6361/201730858},
       adsurl = {https://ui.adsabs.harvard.edu/abs/2017A&A...605A..20C}
}

@ARTICLE{chandrasekhar1953,
       author = {{Chandrasekhar}, S. and {Fermi}, E.},
        title = "{Magnetic Fields in Spiral Arms.}",
      journal = {\apj},
         year = 1953,
        month = jul,
       volume = {118},
        pages = {113},
          doi = {10.1086/145731},
       adsurl = {https://ui.adsabs.harvard.edu/abs/1953ApJ...118..113C}
}

@ARTICLE{ching2016,
       author = {{Ching}, Tao-Chung and {Lai}, Shih-Ping and {Zhang}, Qizhou and {Yang}, Louis and {Girart}, Josep M. and {Rao}, Ramprasad},
        title = "{Helical Magnetic Fields in the NGC 1333 IRAS 4A Protostellar Outflows}",
      journal = {\apj},
         year = 2016,
        month = mar,
       volume = {819},
       number = {2},
          eid = {159},
        pages = {159},
          doi = {10.3847/0004-637X/819/2/159},
archivePrefix = {arXiv},
       eprint = {1601.05229},
 primaryClass = {astro-ph.SR},
       adsurl = {https://ui.adsabs.harvard.edu/abs/2016ApJ...819..159C}
}

@PHDTHESIS{chiong2003,
       author = {{Chiong}, Chau Ching},
        title = "{Zeeman Measurements using SO (1$_0$-0$_1$) Transition and Heterodyne Observations towards the W51 Region}",
       school = {Rheinische Friedrich Wilhelms University of Bonn, Germany},
         year = 2003,
        month = nov,
       adsurl = {https://ui.adsabs.harvard.edu/abs/2003PhDT.......234C}
}

@ARTICLE{ck1983,
       author = {{Crutcher}, R.~M. and {Kazes}, I.},
        title = "{The magnetic field of the NGC 2024 molecular cloud : detection of OH line Zeeman splitting.}",
      journal = {\aap},
         year = 1983,
        month = sep,
       volume = {125},
        pages = {L23-L26},
       adsurl = {https://ui.adsabs.harvard.edu/abs/1983A&A...125L..23C}
}

@ARTICLE{crutcher1996,
       author = {{Crutcher}, Richard M. and {Troland}, Thomas H. and {Lazareff}, Bernard and {Kazes}, Ilya},
        title = "{CN Zeeman Observations of Molecular Cloud Cores}",
      journal = {\apj},
         year = 1996,
        month = jan,
       volume = {456},
        pages = {217},
          doi = {10.1086/176642},
       adsurl = {https://ui.adsabs.harvard.edu/abs/1996ApJ...456..217C}
}

@ARTICLE{crutcher1999c,
       author = {{Crutcher}, Richard M. and {Troland}, Thomas H. and {Lazareff}, Bernard and {Paubert}, Gabriel and {Kaz{\`e}s}, Ilya},
        title = "{Detection of the CN Zeeman Effect in Molecular Clouds}",
      journal = {\apjl},
         year = 1999,
        month = apr,
       volume = {514},
       number = {2},
        pages = {L121-L124},
          doi = {10.1086/311952},
       adsurl = {https://ui.adsabs.harvard.edu/abs/1999ApJ...514L.121C}
}

@ARTICLE{crutcher1993,
       author = {{Crutcher}, R.~M. and {Troland}, T.~H. and {Goodman}, A.~A. and {Heiles}, C. and {Kazes}, I. and {Myers}, P.~C.},
        title = "{OH Zeeman Observations of Dark Clouds}",
      journal = {\apj},
         year = 1993,
        month = apr,
       volume = {407},
        pages = {175},
          doi = {10.1086/172503},
       adsurl = {https://ui.adsabs.harvard.edu/abs/1993ApJ...407..175C}
}

@ARTICLE{chen2025,
       author = {{Chen}, J.~L. and {Zhang}, J.~S. and {Ge}, J.~X. and {Wang}, Y.~X. and {Yu}, H.~Z. and {Zou}, Y.~P. and {Yan}, Y.~T. and {Wang}, X.~Y. and {Wei}, D.~Y.},
        title = "{The Chemical Clock of High-mass Star-forming Regions: N$_{2}$H$^{+}$/CCS}",
      journal = {\aj},
         year = 2025,
        month = aug,
       volume = {170},
       number = {2},
          eid = {74},
        pages = {74},
          doi = {10.3847/1538-3881/addf35},
archivePrefix = {arXiv},
       eprint = {2505.23221},
 primaryClass = {astro-ph.GA},
       adsurl = {https://ui.adsabs.harvard.edu/abs/2025AJ....170...74C}
}

@ARTICLE{crutcher1999a,
       author = {{Crutcher}, R.~M. and {Roberts}, D.~A. and {Troland}, T.~H. and {Goss}, W.~M.},
        title = "{The Magnetic Field of the NGC 2024 Molecular Cloud}",
      journal = {\apj},
         year = 1999,
        month = apr,
       volume = {515},
       number = {1},
        pages = {275-285},
          doi = {10.1086/307026},
       adsurl = {https://ui.adsabs.harvard.edu/abs/1999ApJ...515..275C}
}

@ARTICLE{crutcher1999b,
       author = {{Crutcher}, Richard M.},
        title = "{Magnetic Fields in Molecular Clouds: Observations Confront Theory}",
      journal = {\apj},
         year = 1999,
        month = aug,
       volume = {520},
       number = {2},
        pages = {706-713},
          doi = {10.1086/307483},
       adsurl = {https://ui.adsabs.harvard.edu/abs/1999ApJ...520..706C}
}

@ARTICLE{crutcher2000,
       author = {{Crutcher}, Richard M. and {Troland}, Thomas H.},
        title = "{OH Zeeman Measurement of the Magnetic Field in the L1544 Core}",
      journal = {\apjl},
         year = 2000,
        month = jul,
       volume = {537},
       number = {2},
        pages = {L139-L142},
          doi = {10.1086/312770},
       adsurl = {https://ui.adsabs.harvard.edu/abs/2000ApJ...537L.139C}
}

@ARTICLE{crutcher2004,
       author = {{Crutcher}, Richard M. and {Nutter}, D.~J. and {Ward-Thompson}, D. and {Kirk}, J.~M.},
        title = "{SCUBA Polarization Measurements of the Magnetic Field Strengths in the L183, L1544, and L43 Prestellar Cores}",
      journal = {\apj},
         year = 2004,
        month = jan,
       volume = {600},
       number = {1},
        pages = {279-285},
          doi = {10.1086/379705},
archivePrefix = {arXiv},
       eprint = {astro-ph/0305604},
 primaryClass = {astro-ph},
       adsurl = {https://ui.adsabs.harvard.edu/abs/2004ApJ...600..279C}
}

@ARTICLE{crutcher2010,
       author = {{Crutcher}, Richard M. and {Wandelt}, Benjamin and {Heiles}, Carl and {Falgarone}, Edith and {Troland}, Thomas H.},
        title = "{Magnetic Fields in Interstellar Clouds from Zeeman Observations: Inference of Total Field Strengths by Bayesian Analysis}",
      journal = {\apj},
         year = 2010,
        month = dec,
       volume = {725},
       number = {1},
        pages = {466-479},
          doi = {10.1088/0004-637X/725/1/466},
       adsurl = {https://ui.adsabs.harvard.edu/abs/2010ApJ...725..466C}
}

@ARTICLE{crutcher2012,
       author = {{Crutcher}, Richard M.},
        title = "{Magnetic Fields in Molecular Clouds}",
      journal = {\araa},
         year = 2012,
        month = sep,
       volume = {50},
        pages = {29-63},
          doi = {10.1146/annurev-astro-081811-125514},
       adsurl = {https://ui.adsabs.harvard.edu/abs/2012ARA&A..50...29C}
}

@ARTICLE{crutcher2019,
       author = {{Crutcher}, Richard M. and {Kemball}, Athol J.},
        title = "{Review of Zeeman Effect Observations of Regions of Star Formation K Zeeman Effect, Magnetic Fields, Star formation, Masers, Molecular clouds}",
      journal = {Frontiers in Astronomy and Space Sciences},
         year = 2019,
        month = oct,
       volume = {6},
          eid = {66},
        pages = {66},
          doi = {10.3389/fspas.2019.00066},
archivePrefix = {arXiv},
       eprint = {1911.06210},
 primaryClass = {astro-ph.GA},
       adsurl = {https://ui.adsabs.harvard.edu/abs/2019FrASS...6...66C}
}

@ARTICLE{davis1951,
       author = {{Davis}, Leverett},
        title = "{The Strength of Interstellar Magnetic Fields}",
      journal = {Physical Review},
         year = 1951,
        month = mar,
       volume = {81},
       number = {5},
        pages = {890-891},
          doi = {10.1103/PhysRev.81.890.2},
       adsurl = {https://ui.adsabs.harvard.edu/abs/1951PhRv...81..890D}
}

@ARTICLE{davisgreenstein1951,
       author = {{Davis}, Jr., Leverett and {Greenstein}, Jesse L.},
        title = "{The Polarization of Starlight by Aligned Dust Grains.}",
      journal = {\apj},
         year = 1951,
        month = sep,
       volume = {114},
        pages = {206},
          doi = {10.1086/145464},
       adsurl = {https://ui.adsabs.harvard.edu/abs/1951ApJ...114..206D}
}

@INPROCEEDINGS{Elmegreen1985,
       author = {{Elmegreen}, B.~G.},
        title = "{Molecular clouds and star formation: an overview.}",
    booktitle = {Protostars and Planets II},
         year = 1985,
       editor = {{Black}, D.~C. and {Matthews}, M.~S.},
        month = jan,
        pages = {33-58},
       adsurl = {https://ui.adsabs.harvard.edu/abs/1985prpl.conf...33E}
}

@ARTICLE{Evans1999,
       author = {{Evans}, II, Neal J.},
        title = "{Physical Conditions in Regions of Star Formation}",
      journal = {\araa},
         year = 1999,
        month = jan,
       volume = {37},
        pages = {311-362},
          doi = {10.1146/annurev.astro.37.1.311},
archivePrefix = {arXiv},
       eprint = {astro-ph/9905050},
 primaryClass = {astro-ph},
       adsurl = {https://ui.adsabs.harvard.edu/abs/1999ARA&A..37..311E}
}

@ARTICLE{fiebig1989,
       author = {{Fiebig}, D. and {Guesten}, R.},
        title = "{Strong magnetic fields in interstellar H2O maser clumps.}",
      journal = {\aap},
         year = 1989,
        month = apr,
       volume = {214},
        pages = {333-338},
       adsurl = {https://ui.adsabs.harvard.edu/abs/1989A&A...214..333F}
}

@ARTICLE{fish2005,
       author = {{Fish}, Vincent L. and {Reid}, Mark J. and {Argon}, Alice L. and {Zheng}, Xing-Wu},
        title = "{Full-Polarization Observations of OH Masers in Massive Star-forming Regions. I. Data}",
      journal = {\apjs},
         year = 2005,
        month = sep,
       volume = {160},
       number = {1},
        pages = {220-271},
          doi = {10.1086/431669},
archivePrefix = {arXiv},
       eprint = {astro-ph/0505148},
 primaryClass = {astro-ph},
       adsurl = {https://ui.adsabs.harvard.edu/abs/2005ApJS..160..220F}
}

@ARTICLE{goddi2017,
       author = {{Goddi}, C. and {Surcis}, G. and {Moscadelli}, L. and {Imai}, H. and {Vlemmings}, W.~H.~T. and {van Langevelde}, H.~J. and {Sanna}, A.},
        title = "{Measuring magnetic fields from water masers in the synchrotron protostellar jet in W3(H$_{2}$O)}",
      journal = {\aap},
         year = 2017,
        month = jan,
       volume = {597},
          eid = {A43},
        pages = {A43},
          doi = {10.1051/0004-6361/201629321},
archivePrefix = {arXiv},
       eprint = {1608.02951},
 primaryClass = {astro-ph.GA},
       adsurl = {https://ui.adsabs.harvard.edu/abs/2017A&A...597A..43G}
}

@ARTICLE{goldreich1981,
       author = {{Goldreich}, P. and {Kylafis}, N.~D.},
        title = "{On mapping the magnetic field direction in molecular clouds by polarization measurements}",
      journal = {\apjl},
         year = 1981,
        month = jan,
       volume = {243},
        pages = {L75-L78},
          doi = {10.1086/183446},
       adsurl = {https://ui.adsabs.harvard.edu/abs/1981ApJ...243L..75G}
}

@ARTICLE{greve1980,
       author = {{Greve}, A. and {Pauls}, T.},
        title = "{On the Zeeman splitting of high N recombination lines}",
      journal = {\aap},
         year = 1980,
        month = feb,
       volume = {82},
       number = {3},
        pages = {388},
       adsurl = {https://ui.adsabs.harvard.edu/abs/1980A&A....82..388G}
}

@ARTICLE{gupta2009,
       author = {{Gupta}, H. and {Gottlieb}, C.~A. and {McCarthy}, M.~C. and {Thaddeus}, P.},
        title = "{A Survey of C$_{4}$H, C$_{6}$H, and C$_{6}$H$^{-}$ With the Green Bank Telescope}",
      journal = {\apj},
         year = 2009,
        month = feb,
       volume = {691},
       number = {2},
        pages = {1494-1500},
          doi = {10.1088/0004-637X/691/2/1494},
       adsurl = {https://ui.adsabs.harvard.edu/abs/2009ApJ...691.1494G}
}

@INPROCEEDINGS{guesten1990,
       author = {{Guesten}, Rolf and {Fiebig}, Dirk},
        title = "{Magnetic Fields in Dark Cloud Cores and H2O Masers}",
    booktitle = {Galactic and Intergalactic Magnetic Fields},
         year = 1990,
       editor = {{Beck}, R. and {Kronberg}, P.~P. and {Wielebinski}, R.},
       series = {IAU Symposium},
       volume = {140},
        month = jan,
        pages = {305},
       adsurl = {https://ui.adsabs.harvard.edu/abs/1990IAUS..140..305G}
}

@ARTICLE{guesten1994,
       author = {{Guesten}, R. and {Fiebig}, D. and {Uchida}, K.~I.},
        title = "{The magnetic field in the dense gas bordering W3(OH)}",
      journal = {\aap},
         year = 1994,
        month = jun,
       volume = {286},
        pages = {L51-L54},
       adsurl = {https://ui.adsabs.harvard.edu/abs/1994A&A...286L..51G}
}

@ARTICLE{hansen1982,
       author = {{Hansen}, S.~S.},
        title = "{The magnetic fields in the Orion Kleinmann-Low Nebula as derived from hydroxyl maser radiation}",
      journal = {\apj},
         year = 1982,
        month = sep,
       volume = {260},
        pages = {599-603},
          doi = {10.1086/160281},
       adsurl = {https://ui.adsabs.harvard.edu/abs/1982ApJ...260..599H}
}

@ARTICLE{harvey1974,
       author = {{Harvey}, P.~J. and {Booth}, R.~S. and {Davies}, R.~D. and {Whittet}, D.~C.~B. and {McLaughlin}, W.},
        title = "{Interferometric observations of the structure of main-line OH sources.}",
      journal = {\mnras},
         year = 1974,
        month = dec,
       volume = {169},
        pages = {545-576},
          doi = {10.1093/mnras/169.3.545},
       adsurl = {https://ui.adsabs.harvard.edu/abs/1974MNRAS.169..545H}
}

@INPROCEEDINGS{heiles1993,
       author = {{Heiles}, Carl and {Goodman}, Alyssa A. and {McKee}, Christopher F. and {Zweibel}, Ellen G.},
        title = "{Magnetic Fields in Star-Forming Regions - Observations}",
    booktitle = {Protostars and Planets III},
         year = 1993,
       editor = {{Levy}, Eugene H. and {Lunine}, Jonathan I.},
        month = jan,
        pages = {279},
       adsurl = {https://ui.adsabs.harvard.edu/abs/1993prpl.conf..279H}
}

@ARTICLE{hirota2009,
       author = {{Hirota}, Tomoya and {Ohishi}, Masatoshi and {Yamamoto}, Satoshi},
        title = "{A Search for Carbon-Chain-rich Cores in Dark Clouds}",
      journal = {\apj},
         year = 2009,
        month = jul,
       volume = {699},
       number = {1},
        pages = {585-602},
          doi = {10.1088/0004-637X/699/1/585},
archivePrefix = {arXiv},
       eprint = {0905.3511},
 primaryClass = {astro-ph.SR},
       adsurl = {https://ui.adsabs.harvard.edu/abs/2009ApJ...699..585H}
}

@ARTICLE{hirota2011,
       author = {{Hirota}, Tomoya and {Sakai}, Takeshi and {Sakai}, Nami and {Yamamoto}, Satoshi},
        title = "{Detection of Two Carbon-chain-rich Cores: CB130-3 and L673-SMM4}",
      journal = {\apj},
         year = 2011,
        month = jul,
       volume = {736},
       number = {1},
          eid = {4},
        pages = {4},
          doi = {10.1088/0004-637X/736/1/4},
archivePrefix = {arXiv},
       eprint = {1105.0081},
 primaryClass = {astro-ph.SR},
       adsurl = {https://ui.adsabs.harvard.edu/abs/2011ApJ...736....4H}
}

@INPROCEEDINGS{heiles2009,
       author = {{Heiles}, Carl and {Robishaw}, Timothy},
        title = "{Zeeman splitting in the diffuse interstellar medium-The Milky Way and beyond}",
    booktitle = {Cosmic Magnetic Fields: From Planets, to Stars and Galaxies},
         year = 2009,
       editor = {{Strassmeier}, Klaus G. and {Kosovichev}, Alexander G. and {Beckman}, John E.},
       series = {IAU Symposium},
       volume = {259},
        month = apr,
        pages = {579-590},
          doi = {10.1017/S174392130903141X},
       adsurl = {https://ui.adsabs.harvard.edu/abs/2009IAUS..259..579H}
}

@ARTICLE{klaassen2024,
       author = {{Klaassen}, Pamela and {Traficante}, Alessio and {Beltr{\'a}n}, Maria and {Pattle}, Kate and {Booth}, Mark and {Lovell}, Joshua and {Marshall}, Jonathan and {Hacar}, Alvaro and {Gaches}, Brandt and {Bot}, Caroline and {Peretto}, Nicolas and {Stanke}, Thomas and {Arzoumanian}, Doris and {Duarte Cabral}, Ana and {Duch{\^e}ne}, Gaspard and {Eden}, David and {Hales}, Antonio and {Kauffmann}, Jens and {Luppe}, Patricia and {Marino}, Sebastian and {Redaelli}, Elena and {Rigby}, Andrew and {S{\'a}nchez-Monge}, {\'A}lvaro and {Schisano}, Eugenio and {Semenov}, Dmitry and {Spezzano}, Silvia and {Thompson}, Mark and {Wyrowski}, Friedrich and {Cicone}, Claudia and {Mroczkowski}, Tony and {Cordiner}, Martin and {Di Mascolo}, Luca and {Johnstone}, Doug and {van Kampen}, Eelco and {Lee}, Minju and {Liu}, Daizhong and {Maccarone}, Thomas and {Saintonge}, Am{\'e}lie and {Smith}, Matthew and {Thelen}, Alexander and {Wedemeyer}, Sven},
        title = "{Atacama Large Aperture Submillimeter Telescope (AtLAST) science: Our Galaxy}",
      journal = {Open Research Europe},
         year = 2024,
        month = jun,
       volume = {4},
        pages = {112},
          doi = {10.12688/openreseurope.17450.1},
archivePrefix = {arXiv},
       eprint = {2403.00917},
 primaryClass = {astro-ph.GA},
       adsurl = {https://ui.adsabs.harvard.edu/abs/2024ORE.....4..112K}
}

@ARTICLE{koley2021,
       author = {{Koley}, Atanu and {Roy}, Nirupam and {Menten}, Karl M. and {Jacob}, Arshia M. and {Pillai}, Thushara G.~S. and {Rugel}, Michael R.},
        title = "{The magnetic field in the dense photodissociation region of DR 21}",
      journal = {\mnras},
         year = 2021,
        month = mar,
       volume = {501},
       number = {4},
        pages = {4825-4836},
          doi = {10.1093/mnras/staa3898},
archivePrefix = {arXiv},
       eprint = {2012.08253},
 primaryClass = {astro-ph.GA},
       adsurl = {https://ui.adsabs.harvard.edu/abs/2021MNRAS.501.4825K},
}

@INPROCEEDINGS{beuther2007,
       author = {{Beuther}, H. and {Churchwell}, E.~B. and {McKee}, C.~F. and {Tan}, J.~C.},
        title = "{The Formation of Massive Stars}",
    booktitle = {Protostars and Planets V},
         year = 2007,
       editor = {{Reipurth}, Bo and {Jewitt}, David and {Keil}, Klaus},
        month = jan,
        pages = {165},
          doi = {10.48550/arXiv.astro-ph/0602012},
archivePrefix = {arXiv},
       eprint = {astro-ph/0602012},
 primaryClass = {astro-ph},
       adsurl = {https://ui.adsabs.harvard.edu/abs/2007prpl.conf..165B}
}

@ARTICLE{motte2018,
       author = {{Motte}, Fr{\'e}d{\'e}rique and {Bontemps}, Sylvain and {Louvet}, Fabien},
        title = "{High-Mass Star and Massive Cluster Formation in the Milky Way}",
      journal = {\araa},
         year = 2018,
        month = sep,
       volume = {56},
        pages = {41-82},
          doi = {10.1146/annurev-astro-091916-055235},
archivePrefix = {arXiv},
       eprint = {1706.00118},
 primaryClass = {astro-ph.GA},
       adsurl = {https://ui.adsabs.harvard.edu/abs/2018ARA&A..56...41M}
}

@ARTICLE{beuther2025,
       author = {{Beuther}, H. and {Kuiper}, R. and {Tafalla}, M.},
        title = "{Star Formation from Low to High Mass: A Comparative View}",
      journal = {\araa},
         year = 2025,
        month = aug,
       volume = {63},
       number = {1},
        pages = {1-44},
          doi = {10.1146/annurev-astro-013125-122023},
archivePrefix = {arXiv},
       eprint = {2501.16866},
 primaryClass = {astro-ph.GA},
       adsurl = {https://ui.adsabs.harvard.edu/abs/2025ARA&A..63....1B}
}

@ARTICLE{worthen2025,
       author = {{Worthen}, Kadin and {Svoboda}, Brian E. and {Meier}, David S. and {Ott}, Juergen and {Friesen}, Rachel and {Patience}, Jennifer and {Shirley}, Yancy},
        title = "{The Young Ages of 70 {\ensuremath{\mu}}m Dark Clumps Inferred from Carbon Chain Chemistry}",
      journal = {\apj},
         year = 2025,
        month = mar,
       volume = {981},
       number = {2},
          eid = {207},
        pages = {207},
          doi = {10.3847/1538-4357/adabdd},
archivePrefix = {arXiv},
       eprint = {2502.04283},
 primaryClass = {astro-ph.GA},
       adsurl = {https://ui.adsabs.harvard.edu/abs/2025ApJ...981..207W}
}

@INPROCEEDINGS{urquhart2024,
       author = {{Urquhart}, J.~S.},
        title = "{Evolutionary Trends in Star Formation}",
    booktitle = {Cosmic Masers: Proper Motion Toward the Next-Generation Large Projects},
         year = 2024,
       editor = {{Hirota}, Tomoya and {Imai}, Hiroshi and {Menten}, Karl and {Pihlstr{\"o}m}, Ylva},
       series = {IAU Symposium},
       volume = {380},
        month = jan,
        pages = {135-151},
          doi = {10.1017/S1743921323002326},
       adsurl = {https://ui.adsabs.harvard.edu/abs/2024IAUS..380..135U}
}

@ARTICLE{han2017,
       author = {{Han}, J.~L.},
        title = "{Observing Interstellar and Intergalactic Magnetic Fields}",
      journal = {\araa},
         year = 2017,
        month = aug,
       volume = {55},
       number = {1},
        pages = {111-157},
          doi = {10.1146/annurev-astro-091916-055221},
       adsurl = {https://ui.adsabs.harvard.edu/abs/2017ARA&A..55..111H}
}

@ARTICLE{koley2022,
       author = {{Koley}, Atanu and {Roy}, Nirupam and {Momjian}, Emmanuel and {Sarma}, Anuj P. and {Datta}, Abhirup},
        title = "{Magnetic field measurement in TMC-1C using 22.3 GHz CCS Zeeman splitting}",
      journal = {\mnras},
         year = 2022,
        month = oct,
       volume = {516},
       number = {1},
        pages = {L48-L52},
          doi = {10.1093/mnrasl/slac085},
archivePrefix = {arXiv},
       eprint = {2207.12604},
 primaryClass = {astro-ph.GA},
       adsurl = {https://ui.adsabs.harvard.edu/abs/2022MNRAS.516L..48K}
}

@ARTICLE{kaifu2004,
       author = {{Kaifu}, Norio and {Ohishi}, Masatoshi and {Kawaguchi}, Kentarou and {Saito}, Shuji and {Yamamoto}, Satoshi and {Miyaji}, Takeshi and {Miyazawa}, Keisuke and {Ishikawa}, Shin-Ichi and {Noumaru}, Chiaki and {Harasawa}, Sumiko and {Okuda}, Michiko and {Suzuki}, Hiroko},
        title = "{A 8.8--50GHz Complete Spectral Line Survey toward TMC-1 I. Survey Data}",
      journal = {\pasj},
         year = 2004,
        month = feb,
       volume = {56},
        pages = {69-173},
          doi = {10.1093/pasj/56.1.69},
       adsurl = {https://ui.adsabs.harvard.edu/abs/2004PASJ...56...69K}
}

@ARTICLE{lankhaar2018,
       author = {{Lankhaar}, Boy and {Vlemmings}, Wouter and {Surcis}, Gabriele and {van Langevelde}, Huib Jan and {Groenenboom}, Gerrit C. and {van der Avoird}, Ad},
        title = "{Characterization of methanol as a magnetic field tracer in star-forming regions}",
      journal = {Nature Astronomy},
         year = 2018,
        month = feb,
       volume = {2},
        pages = {145-150},
          doi = {10.1038/s41550-017-0341-8},
archivePrefix = {arXiv},
       eprint = {1802.05764},
 primaryClass = {astro-ph.GA},
       adsurl = {https://ui.adsabs.harvard.edu/abs/2018NatAs...2..145L}
}

@ARTICLE{larsson2019,
       author = {{Larsson}, Richard and {Lankhaar}, Boy and {Eriksson}, Patrick},
        title = "{Updated Zeeman effect splitting coefficients for molecular oxygen in planetary applications}",
      journal = {\jqsrt},
         year = 2019,
        month = feb,
       volume = {224},
        pages = {431-438},
          doi = {10.1016/j.jqsrt.2018.12.004},
       adsurl = {https://ui.adsabs.harvard.edu/abs/2019JQSRT.224..431L}
}

@ARTICLE{larsson2020,
       author = {{Larsson}, Richard and {Lankhaar}, Boy},
        title = "{Zeeman effect splitting coefficients for ClO, OH and NO in some earth atmosphere applications}",
      journal = {\jqsrt},
         year = 2020,
        month = jul,
       volume = {250},
          eid = {107050},
        pages = {107050},
          doi = {10.1016/j.jqsrt.2020.107050},
       adsurl = {https://ui.adsabs.harvard.edu/abs/2020JQSRT.25007050L}
}

@ARTICLE{lis2025,
       author = {{Lis}, Dariusz C. and {Langer}, William D. and {Pineda}, Jorge L. and {Gandhi}, Kahaan and {Willacy}, Karen and {Goldsmith}, Paul F. and {Widicus Weaver}, Susanna and {Majumdar}, Liton and {Seo}, Youngmin and {Horiuchi}, Shinji and {Bop}, Cheikh T. and {Lique}, Fran{\c{c}}ois},
        title = "{An 18{\textendash}25 GHz spectroscopic survey of dense cores in the Chamaeleon I molecular cloud}",
      journal = {\aap},
         year = 2025,
        month = apr,
       volume = {696},
          eid = {A61},
        pages = {A61},
          doi = {10.1051/0004-6361/202553856},
archivePrefix = {arXiv},
       eprint = {2503.01755},
 primaryClass = {astro-ph.GA},
       adsurl = {https://ui.adsabs.harvard.edu/abs/2025A&A...696A..61L}
}

@ARTICLE{li2015,
       author = {{Li}, Juan and {Wang}, Junzhi and {Zhu}, Qingfeng and {Zhang}, Jiangshui and {Li}, Di},
        title = "{Sulfur-bearing Molecules in Massive Star-forming Regions: Observations of OCS, CS, H$_{2}$S, and SO}",
      journal = {\apj},
         year = 2015,
        month = mar,
       volume = {802},
       number = {1},
          eid = {40},
        pages = {40},
          doi = {10.1088/0004-637X/802/1/40},
archivePrefix = {arXiv},
       eprint = {1501.06018},
 primaryClass = {astro-ph.SR},
       adsurl = {https://ui.adsabs.harvard.edu/abs/2015ApJ...802...40L}
}

@ARTICLE{zhao2024,
       author = {{Zhao}, Mengke and {Li}, Guang-Xing and {Qiu}, Keping},
        title = "{Slope of Magnetic Field{\textendash}Density Relation as an Indicator of Magnetic Dominance}",
      journal = {\apj},
         year = 2024,
        month = dec,
       volume = {976},
       number = {2},
          eid = {209},
        pages = {209},
          doi = {10.3847/1538-4357/ad8b4d},
archivePrefix = {arXiv},
       eprint = {2409.02786},
 primaryClass = {astro-ph.GA},
       adsurl = {https://ui.adsabs.harvard.edu/abs/2024ApJ...976..209Z}
}

@ARTICLE{liu2022,
       author = {{Liu}, Junhao and {Zhang}, Qizhou and {Qiu}, Keping},
        title = "{Magnetic field properties in star formation: A review of their analysis methods and interpretation}",
      journal = {Frontiers in Astronomy and Space Sciences},
         year = 2022,
        month = sep,
       volume = {9},
          eid = {943556},
        pages = {943556},
          doi = {10.3389/fspas.2022.943556},
archivePrefix = {arXiv},
       eprint = {2208.06492},
 primaryClass = {astro-ph.GA},
       adsurl = {https://ui.adsabs.harvard.edu/abs/2022FrASS...9.3556L}
}

@ARTICLE{jiang2020,
       author = {{Jiang}, Hangjin and {Li}, Hua-bai and {Fan}, Xiaodan},
        title = "{Bayesian Revisit of the Relationship between the Total Field Strength and the Volume Density of Interstellar Clouds}",
      journal = {\apj},
         year = 2020,
        month = feb,
       volume = {890},
       number = {2},
          eid = {153},
        pages = {153},
          doi = {10.3847/1538-4357/ab672b},
       adsurl = {https://ui.adsabs.harvard.edu/abs/2020ApJ...890..153J}
}

@INPROCEEDINGS{zinchenko2018,
       author = {{Zinchenko}, Igor and {Henkel}, Christian},
        title = "{SO survey of massive cores}",
    booktitle = {Astrochemistry VII: Through the Cosmos from Galaxies to Planets},
         year = 2018,
       editor = {{Cunningham}, Maria and {Millar}, Tom and {Aikawa}, Yuri},
       series = {IAU Symposium},
       volume = {332},
        month = sep,
        pages = {274-277},
          doi = {10.1017/S1743921317007694},
archivePrefix = {arXiv},
       eprint = {1711.09380},
 primaryClass = {astro-ph.GA},
       adsurl = {https://ui.adsabs.harvard.edu/abs/2018IAUS..332..274Z}
}

@ARTICLE{codella1997,
       author = {{Codella}, C. and {Muders}, D.},
        title = "{SO observations towards BOK globules}",
      journal = {\mnras},
         year = 1997,
        month = oct,
       volume = {291},
       number = {2},
        pages = {337-344},
          doi = {10.1093/mnras/291.2.337},
       adsurl = {https://ui.adsabs.harvard.edu/abs/1997MNRAS.291..337C}
}

@ARTICLE{rydbeck1980,
       author = {{Rydbeck}, O.~E.~H. and {Hjalmarson}, A. and {Rydbeck}, G. and {Ellder}, J. and {Kollberg}, E. and {Irvine}, W.~M.},
        title = "{Observations of SO in dark and molecular clouds}",
      journal = {\apjl},
         year = 1980,
        month = feb,
       volume = {235},
        pages = {L171-L175},
          doi = {10.1086/183184},
       adsurl = {https://ui.adsabs.harvard.edu/abs/1980ApJ...235L.171R}
}

@ARTICLE{foster2009,
       author = {{Foster}, Jonathan B. and {Rosolowsky}, Erik W. and {Kauffmann}, Jens and {Pineda}, Jaime E. and {Borkin}, Michelle A. and {Caselli}, Paola and {Myers}, Phil C. and {Goodman}, Alyssa A.},
        title = "{Dense Cores in Perseus: The Influence of Stellar Content and Cluster Environment}",
      journal = {\apj},
         year = 2009,
        month = may,
       volume = {696},
       number = {1},
        pages = {298-319},
          doi = {10.1088/0004-637X/696/1/298},
archivePrefix = {arXiv},
       eprint = {0902.0536},
 primaryClass = {astro-ph.SR},
       adsurl = {https://ui.adsabs.harvard.edu/abs/2009ApJ...696..298F}
}

@ARTICLE{marka2012,
       author = {{Marka}, C. and {Schreyer}, K. and {Launhardt}, R. and {Semenov}, D.~A. and {Henning}, Th.},
        title = "{Tracing the evolutionary stage of Bok globules: CCS and NH$_{3}$}",
      journal = {\aap},
         year = 2012,
        month = jan,
       volume = {537},
          eid = {A4},
        pages = {A4},
          doi = {10.1051/0004-6361/201014375},
archivePrefix = {arXiv},
       eprint = {1110.3945},
 primaryClass = {astro-ph.GA},
       adsurl = {https://ui.adsabs.harvard.edu/abs/2012A&A...537A...4M}
}

@ARTICLE{li2021,
       author = {{Li}, Hua-Bai},
        title = "{Magnetic Fields in Molecular Clouds{\textemdash}Observation and Interpretation}",
      journal = {Galaxies},
         year = 2021,
        month = jun,
       volume = {9},
       number = {2},
          eid = {41},
        pages = {41},
          doi = {10.3390/galaxies9020041},
archivePrefix = {arXiv},
       eprint = {2106.08172},
 primaryClass = {astro-ph.GA},
       adsurl = {https://ui.adsabs.harvard.edu/abs/2021Galax...9...41L}
}

@ARTICLE{liu2020,
       author = {{Liu}, Tie and {Evans}, Neal J. and {Kim}, Kee-Tae and {Goldsmith}, Paul F. and {Liu}, Sheng-Yuan and {Zhang}, Qizhou and {Tatematsu}, Ken'ichi and {Wang}, Ke and {Juvela}, Mika and {Bronfman}, Leonardo and {Cunningham}, Maria R. and {Garay}, Guido and {Hirota}, Tomoya and {Lee}, Jeong-Eun and {Kang}, Sung-Ju and {Li}, Di and {Li}, Pak-Shing and {Mardones}, Diego and {Qin}, Sheng-Li and {Ristorcelli}, Isabelle and {Tej}, Anandmayee and {Toth}, L. Viktor and {Wu}, Jing-Wen and {Wu}, Yue-Fang and {Yi}, Hee-weon and {Yun}, Hyeong-Sik and {Liu}, Hong-Li and {Peng}, Ya-Ping and {Li}, Juan and {Li}, Shang-Huo and {Lee}, Chang Won and {Shen}, Zhi-Qiang and {Baug}, Tapas and {Wang}, Jun-Zhi and {Zhang}, Yong and {Issac}, Namitha and {Zhu}, Feng-Yao and {Luo}, Qiu-Yi and {Soam}, Archana and {Liu}, Xun-Chuan and {Xu}, Feng-Wei and {Wang}, Yu and {Zhang}, Chao and {Ren}, Zhiyuan and {Zhang}, Chao},
        title = "{ATOMS: ALMA Three-millimeter Observations of Massive Star-forming regions - I. Survey description and a first look at G9.62+0.19}",
      journal = {\mnras},
         year = 2020,
        month = aug,
       volume = {496},
       number = {3},
        pages = {2790-2820},
          doi = {10.1093/mnras/staa1577},
archivePrefix = {arXiv},
       eprint = {2006.01549},
 primaryClass = {astro-ph.GA},
       adsurl = {https://ui.adsabs.harvard.edu/abs/2020MNRAS.496.2790L}
}

@ARTICLE{levin2001,
       author = {{Levin}, S.~M. and {Langer}, W.~D. and {Velusamy}, T. and {Kuiper}, T.~B.~H. and {Crutcher}, R.~M.},
        title = "{Measuring the Magnetic Field Strength in L1498 with Zeeman-splitting Observations of CCS}",
      journal = {\apj},
         year = 2001,
        month = jul,
       volume = {555},
       number = {2},
        pages = {850-854},
          doi = {10.1086/321518},
       adsurl = {https://ui.adsabs.harvard.edu/abs/2001ApJ...555..850L}
}

@ARTICLE{matthews1986,
       author = {{Matthews}, H.~E. and {Bell}, M.~B. and {Sears}, T.~J. and {Turner}, B.~E. and {Rickard}, L.~J.},
        title = "{A search for rotationally-excited CH in galactic sources.}",
      journal = {\aap},
         year = 1986,
        month = jun,
       volume = {161},
        pages = {329-333},
       adsurl = {https://ui.adsabs.harvard.edu/abs/1986A&A...161..329M}
}

@ARTICLE{mestel1966,
       author = {{Mestel}, L.},
        title = "{The magnetic field of a contracting gas cloud. I,Strict flux-freezing}",
      journal = {\mnras},
         year = 1966,
        month = jan,
       volume = {133},
        pages = {265},
          doi = {10.1093/mnras/133.2.265},
       adsurl = {https://ui.adsabs.harvard.edu/abs/1966MNRAS.133..265M}
}

@ARTICLE{momjian2012,
       author = {{Momjian}, E. and {Sarma}, A.~P.},
        title = "{Comparison of Two Epochs of the Zeeman Effect in the 44 GHz Class I Methanol (CH$_{3}$OH) Maser Line in OMC-2}",
      journal = {\aj},
         year = 2012,
        month = dec,
       volume = {144},
       number = {6},
          eid = {189},
        pages = {189},
          doi = {10.1088/0004-6256/144/6/189},
archivePrefix = {arXiv},
       eprint = {1210.5951},
 primaryClass = {astro-ph.GA},
       adsurl = {https://ui.adsabs.harvard.edu/abs/2012AJ....144..189M}
}

@ARTICLE{momjian2017,
       author = {{Momjian}, E. and {Sarma}, A.~P.},
        title = "{The Zeeman Effect in the 44 GHz Class I Methanol Maser Line toward DR21(OH)}",
      journal = {\apj},
         year = 2017,
        month = jan,
       volume = {834},
       number = {2},
          eid = {168},
        pages = {168},
          doi = {10.3847/1538-4357/834/2/168},
archivePrefix = {arXiv},
       eprint = {1610.09047},
 primaryClass = {astro-ph.GA},
       adsurl = {https://ui.adsabs.harvard.edu/abs/2017ApJ...834..168M}
}

@ARTICLE{momjian2019,
       author = {{Momjian}, E. and {Sarma}, A.~P.},
        title = "{The Zeeman Effect in the 44 GHz Class I Methanol (CH$_{3}$OH) Maser Line toward DR21W}",
      journal = {\apj},
         year = 2019,
        month = feb,
       volume = {872},
       number = {1},
          eid = {12},
        pages = {12},
          doi = {10.3847/1538-4357/aafad8},
archivePrefix = {arXiv},
       eprint = {1812.09977},
 primaryClass = {astro-ph.GA},
       adsurl = {https://ui.adsabs.harvard.edu/abs/2019ApJ...872...12M}
}

@ARTICLE{moran1978,
       author = {{Moran}, J.~M. and {Reid}, M.~J. and {Lada}, C.~J. and {Yen}, J.~L. and {Johnston}, K.~J. and {Spencer}, J.~H.},
        title = "{Evidence for the Zeeman effect in the OH maser emission from W3(OH).}",
      journal = {\apjl},
         year = 1978,
        month = sep,
       volume = {224},
        pages = {L67-L71},
          doi = {10.1086/182761},
       adsurl = {https://ui.adsabs.harvard.edu/abs/1978ApJ...224L..67M}
}

@INPROCEEDINGS{mouschovias1999,
       author = {{Mouschovias}, Telemachos Ch. and {Ciolek}, Glenn E.},
        title = "{Magnetic Fields and Star Formation: A Theory Reaching Adulthood}",
    booktitle = {The Origin of Stars and Planetary Systems},
         year = 1999,
       editor = {{Lada}, Charles J. and {Kylafis}, Nikolaos D.},
       series = {NATO Science Series},
       volume = {540},
        month = jan,
        pages = {305},
       adsurl = {https://ui.adsabs.harvard.edu/abs/1999ASIC..540..305M}
}

@ARTICLE{mouschovias2010,
       author = {{Mouschovias}, Telemachos Ch. and {Tassis}, Konstantinos},
        title = "{Self-consistent analysis of OH-Zeeman observations: too much noise about noise}",
      journal = {\mnras},
         year = 2010,
        month = dec,
       volume = {409},
       number = {2},
        pages = {801-807},
          doi = {10.1111/j.1365-2966.2010.17345.x},
archivePrefix = {arXiv},
       eprint = {1007.3741},
 primaryClass = {astro-ph.GA},
       adsurl = {https://ui.adsabs.harvard.edu/abs/2010MNRAS.409..801M}
}

@ARTICLE{mazzei2020,
       author = {{Mazzei}, Renato and {Cleeves}, L. Ilsedore and {Li}, Zhi-Yun},
        title = "{Untangling Magnetic Complexity in Protoplanetary Disks with the Zeeman Effect}",
      journal = {\apj},
         year = 2020,
        month = nov,
       volume = {903},
       number = {1},
          eid = {20},
        pages = {20},
          doi = {10.3847/1538-4357/abb67a},
archivePrefix = {arXiv},
       eprint = {2009.08996},
 primaryClass = {astro-ph.SR},
       adsurl = {https://ui.adsabs.harvard.edu/abs/2020ApJ...903...20M}
}

@ARTICLE{vlemmings2019,
       author = {{Vlemmings}, W.~H.~T. and {Lankhaar}, B. and {Cazzoletti}, P. and {Ceccobello}, C. and {Dall'Olio}, D. and {van Dishoeck}, E.~F. and {Facchini}, S. and {Humphreys}, E.~M.~L. and {Persson}, M.~V. and {Testi}, L. and {Williams}, J.~P.},
        title = "{Stringent limits on the magnetic field strength in the disc of TW Hya. ALMA observations of CN polarisation}",
      journal = {\aap},
         year = 2019,
        month = apr,
       volume = {624},
          eid = {L7},
        pages = {L7},
          doi = {10.1051/0004-6361/201935459},
archivePrefix = {arXiv},
       eprint = {1904.01632},
 primaryClass = {astro-ph.SR},
       adsurl = {https://ui.adsabs.harvard.edu/abs/2019A&A...624L...7V}
}

@ARTICLE{churchwell2002,
       author = {{Churchwell}, Ed},
        title = "{Ultra-Compact HII Regions and Massive Star Formation}",
      journal = {\araa},
         year = 2002,
        month = jan,
       volume = {40},
        pages = {27-62},
          doi = {10.1146/annurev.astro.40.060401.093845},
       adsurl = {https://ui.adsabs.harvard.edu/abs/2002ARA&A..40...27C}
}

@ARTICLE{breen2010,
       author = {{Breen}, S.~L. and {Ellingsen}, S.~P. and {Caswell}, J.~L. and {Lewis}, B.~E.},
        title = "{12.2-GHz methanol masers towards 1.2-mm dust clumps: quantifying high-mass star formation evolutionary schemes}",
      journal = {\mnras},
         year = 2010,
        month = feb,
       volume = {401},
       number = {4},
        pages = {2219-2244},
          doi = {10.1111/j.1365-2966.2009.15831.x},
archivePrefix = {arXiv},
       eprint = {0910.1223},
 primaryClass = {astro-ph.SR},
       adsurl = {https://ui.adsabs.harvard.edu/abs/2010MNRAS.401.2219B}
}

@ARTICLE{mroczkowski2025,
       author = {{Mroczkowski}, Tony and {Gallardo}, Patricio A. and {Timpe}, Martin and {Kiselev}, Aleksej and {Groh}, Manuel and {Kaercher}, Hans and {Reichert}, Matthias and {Cicone}, Claudia and {Puddu}, Roberto and {Dubois-dit-Bonclaude}, Pierre and {Bok}, Daniel and {Dahl}, Erik and {Macintosh}, Mike and {Dicker}, Simon and {Viole}, Isabelle and {Sartori}, Sabrina and {Valenzuela Venegas}, Guillermo Andr{\'e}s and {Zeyringer}, Marianne and {Niemack}, Michael and {Poppi}, Sergio and {Olguin}, Rodrigo and {Hatziminaoglou}, Evanthia and {De Breuck}, Carlos and {Klaassen}, Pamela and {Montenegro-Montes}, Francisco Miguel and {Zimmerer}, Thomas},
        title = "{The conceptual design of the 50-meter Atacama Large Aperture Submillimeter Telescope (AtLAST)}",
      journal = {\aap},
         year = 2025,
        month = feb,
       volume = {694},
          eid = {A142},
        pages = {A142},
          doi = {10.1051/0004-6361/202449786},
archivePrefix = {arXiv},
       eprint = {2402.18645},
 primaryClass = {astro-ph.IM},
       adsurl = {https://ui.adsabs.harvard.edu/abs/2025A&A...694A.142M}
}

@ARTICLE{nakamura2019,
       author = {{Nakamura}, Fumitaka and {Kameno}, Seiji and {Kusune}, Takayoshi and {Mizuno}, Izumi and {Dobashi}, Kazuhito and {Shimoikura}, Tomomi and {Taniguchi}, Kotomi},
        title = "{First clear detection of the CCS Zeeman splitting toward the pre-stellar core, Taurus Molecular Cloud 1}",
      journal = {\pasj},
         year = 2019,
        month = dec,
       volume = {71},
       number = {6},
          eid = {117},
        pages = {117},
          doi = {10.1093/pasj/psz102},
archivePrefix = {arXiv},
       eprint = {1908.07708},
 primaryClass = {astro-ph.SR},
       adsurl = {https://ui.adsabs.harvard.edu/abs/2019PASJ...71..117N}
}

@ARTICLE{nedoluha1992,
       author = {{Nedoluha}, Gerald E. and {Watson}, William D.},
        title = "{The Zeeman Effect in Astrophysical Water Masers and the Observation of Strong Magnetic Fields in Regions of Star Formation}",
      journal = {\apj},
         year = 1992,
        month = jan,
       volume = {384},
        pages = {185},
          doi = {10.1086/170862},
       adsurl = {https://ui.adsabs.harvard.edu/abs/1992ApJ...384..185N}
}

@ARTICLE{oyama2020,
       author = {{Oyama}, Takahiro and {Ozaki}, Hironori and {Sumiyoshi}, Yoshihiro and {Araki}, Mitsunori and {Takano}, Shuro and {Kuze}, Nobuhiko and {Tsukiyama}, Koichi},
        title = "{Reevaluation of C$_{4}$H Abundance Based on the Revised Dipole Moment}",
      journal = {\apj},
         year = 2020,
        month = feb,
       volume = {890},
       number = {1},
          eid = {39},
        pages = {39},
          doi = {10.3847/1538-4357/ab6a0a},
       adsurl = {https://ui.adsabs.harvard.edu/abs/2020ApJ...890...39O}
}

@INPROCEEDINGS{oonk2015,
       author = {{Oonk}, R. and {Morabito}, L. and {Salgado}, F. and {Toribio}, M.~C. and {van Weeren}, R.~J. and {Tielens}, A.~G.~G.~M. and {Rottgering}, H.~J.~A.},
        title = "{The Physics of the Cold Neutral Medium: Low-frequency Radio Recombination Lines with the Square Kilometre Array}",
    booktitle = {Advancing Astrophysics with the Square Kilometre Array (AASKA14)},
         year = 2015,
        month = apr,
          eid = {139},
        pages = {139},
          doi = {10.22323/1.215.0139},
archivePrefix = {arXiv},
       eprint = {1501.01179},
 primaryClass = {astro-ph.GA},
       adsurl = {https://ui.adsabs.harvard.edu/abs/2015aska.confE.139O}
}

@ARTICLE{pattle2017,
       author = {{Pattle}, Kate and {Ward-Thompson}, Derek and {Berry}, David and {Hatchell}, Jennifer and {Chen}, Huei-Ru and {Pon}, Andy and {Koch}, Patrick M. and {Kwon}, Woojin and {Kim}, Jongsoo and {Bastien}, Pierre and {Cho}, Jungyeon and {Coud{\'e}}, Simon and {Di Francesco}, James and {Fuller}, Gary and {Furuya}, Ray S. and {Graves}, Sarah F. and {Johnstone}, Doug and {Kirk}, Jason and {Kwon}, Jungmi and {Lee}, Chang Won and {Matthews}, Brenda C. and {Mottram}, Joseph C. and {Parsons}, Harriet and {Sadavoy}, Sarah and {Shinnaga}, Hiroko and {Soam}, Archana and {Hasegawa}, Tetsuo and {Lai}, Shih-Ping and {Qiu}, Keping and {Friberg}, Per},
        title = "{The JCMT BISTRO Survey: The Magnetic Field Strength in the Orion A Filament}",
      journal = {\apj},
         year = 2017,
        month = sep,
       volume = {846},
       number = {2},
          eid = {122},
        pages = {122},
          doi = {10.3847/1538-4357/aa80e5},
archivePrefix = {arXiv},
       eprint = {1707.05269},
 primaryClass = {astro-ph.GA},
       adsurl = {https://ui.adsabs.harvard.edu/abs/2017ApJ...846..122P}
}

@INPROCEEDINGS{pattle2023,
       author = {{Pattle}, K. and {Fissel}, L. and {Tahani}, M. and {Liu}, T. and {Ntormousi}, E.},
        title = "{Magnetic Fields in Star Formation: from Clouds to Cores}",
    booktitle = {Protostars and Planets VII},
         year = 2023,
       editor = {{Inutsuka}, S. and {Aikawa}, Y. and {Muto}, T. and {Tomida}, K. and {Tamura}, M.},
       series = {ASP Conf. Ser.},
       volume = {534},
        month = jul,
        pages = {193},
          doi = {10.48550/arXiv.2203.11179},
archivePrefix = {arXiv},
       eprint = {2203.11179},
 primaryClass = {astro-ph.GA},
       adsurl = {https://ui.adsabs.harvard.edu/abs/2023ASPC..534..193P}
}

@ARTICLE{pattle2025,
       author = {{Pattle}, Kate and {Karoly}, Janik and {Findlay}, Lorna Buhil and {Coud{\'e}}, Simon and {Hensley}, Brandon S. and {Cortes}, Paulo C. and {Di Francesco}, James and {Le Gouellec}, Valentin J.~M. and {Lopez-Rodriguez}, Enrique and {Louvet}, Fabien},
        title = "{Mapping magnetic fields from clouds to cores with PRIMAger}",
      journal = {Journal of Astronomical Telescopes, Instruments, and Systems},
         year = 2025,
        month = jul,
       volume = {11},
          eid = {031615},
        pages = {031615},
          doi = {10.1117/1.JATIS.11.3.031615},
archivePrefix = {arXiv},
       eprint = {2509.01796},
 primaryClass = {astro-ph.GA},
       adsurl = {https://ui.adsabs.harvard.edu/abs/2025JATIS..11c1615P}
}

@ARTICLE{pineau1993,
       author = {{Pineau des Forets}, G. and {Roueff}, E. and {Schilke}, P. and {Flower}, D.~R.},
        title = "{Sulphur-bearing molecules as tracers of shocks in interstellar clouds}",
      journal = {\mnras},
         year = 1993,
        month = jun,
       volume = {262},
       number = {4},
        pages = {915-928},
          doi = {10.1093/mnras/262.4.915},
       adsurl = {https://ui.adsabs.harvard.edu/abs/1993MNRAS.262..915P}
}

@ARTICLE{pineda2025,
       author = {{Pineda}, Jaime E. and {Friesen}, Rachel K. and {(co-PIs)} and {Rosolowsky}, Erik and {Chac{\'o}n-Tanarro}, Ana and {Chen}, Michael Chun-Yuan and {Di Francesco}, James and {Kirk}, Helen and {Punanova}, Anna and {Seo}, Youngmin and {Shirley}, Yancy and {Ginsburg}, Adam and {Offner}, Stella S.~R. and {Pandhi}, Ayush and {Singh}, Ayushi and {Quan}, Feiyu and {Arce}, H{\'e}ctor G. and {Caselli}, Paola and {Choudhury}, Spandan and {Goodman}, Alyssa A. and {Heitsch}, Fabian and {Martin}, Peter G. and {Matzner}, Christopher D. and {Myers}, Philip C. and {Redaelli}, Elena and {Scibelli}, Samantha and {GAS Collaboration}},
        title = "{The Green Bank Ammonia Survey: Data Release 2}",
      journal = {\apjs},
         year = 2026,
        month = jan,
       volume = {282},
       number = {1},
          eid = {18},
        pages = {18},
          doi = {10.3847/1538-4365/ae11b1},
archivePrefix = {arXiv},
       eprint = {2510.10607},
 primaryClass = {astro-ph.GA},
       adsurl = {https://ui.adsabs.harvard.edu/abs/2026ApJS..282...18P}
}

@ARTICLE{soler2013,
       author = {{Soler}, J.~D. and {Hennebelle}, P. and {Martin}, P.~G. and {Miville-Desch{\^e}nes}, M.-A. and {Netterfield}, C.~B. and {Fissel}, L.~M.},
        title = "{An Imprint of Molecular Cloud Magnetization in the Morphology of the Dust Polarized Emission}",
      journal = {\apj},
         year = 2013,
        month = sep,
       volume = {774},
       number = {2},
          eid = {128},
        pages = {128},
          doi = {10.1088/0004-637X/774/2/128},
archivePrefix = {arXiv},
       eprint = {1303.1830},
 primaryClass = {astro-ph.GA},
       adsurl = {https://ui.adsabs.harvard.edu/abs/2013ApJ...774..128S}
}

@ARTICLE{planckxxxv,
       author = {{Planck Collaboration} and {Ade}, P.~A.~R. and {Aghanim}, N. and {Alves}, M.~I.~R. and {Arnaud}, M. and {Arzoumanian}, D. and {Ashdown}, M. and {Aumont}, J. and {Baccigalupi}, C. and {Banday}, A.~J. and {Barreiro}, R.~B. and {Bartolo}, N. and {Battaner}, E. and {Benabed}, K. and {Beno{\^\i}t}, A. and {Benoit-L{\'e}vy}, A. and {Bernard}, J.-P. and {Bersanelli}, M. and {Bielewicz}, P. and {Bock}, J.~J. and {Bonavera}, L. and {Bond}, J.~R. and {Borrill}, J. and {Bouchet}, F.~R. and {Boulanger}, F. and {Bracco}, A. and {Burigana}, C. and {Calabrese}, E. and {Cardoso}, J.-F. and {Catalano}, A. and {Chiang}, H.~C. and {Christensen}, P.~R. and {Colombo}, L.~P.~L. and {Combet}, C. and {Couchot}, F. and {Crill}, B.~P. and {Curto}, A. and {Cuttaia}, F. and {Danese}, L. and {Davies}, R.~D. and {Davis}, R.~J. and {de Bernardis}, P. and {de Rosa}, A. and {de Zotti}, G. and {Delabrouille}, J. and {Dickinson}, C. and {Diego}, J.~M. and {Dole}, H. and {Donzelli}, S. and {Dor{\'e}}, O. and {Douspis}, M. and {Ducout}, A. and {Dupac}, X. and {Efstathiou}, G. and {Elsner}, F. and {En{\ss}lin}, T.~A. and {Eriksen}, H.~K. and {Falceta-Gon{\c{c}}alves}, D. and {Falgarone}, E. and {Ferri{\`e}re}, K. and {Finelli}, F. and {Forni}, O. and {Frailis}, M. and {Fraisse}, A.~A. and {Franceschi}, E. and {Frejsel}, A. and {Galeotta}, S. and {Galli}, S. and {Ganga}, K. and {Ghosh}, T. and {Giard}, M. and {Gjerl{\o}w}, E. and {Gonz{\'a}lez-Nuevo}, J. and {G{\'o}rski}, K.~M. and {Gregorio}, A. and {Gruppuso}, A. and {Gudmundsson}, J.~E. and {Guillet}, V. and {Harrison}, D.~L. and {Helou}, G. and {Hennebelle}, P. and {Henrot-Versill{\'e}}, S. and {Hern{\'a}ndez-Monteagudo}, C. and {Herranz}, D. and {Hildebrandt}, S.~R. and {Hivon}, E. and {Holmes}, W.~A. and {Hornstrup}, A. and {Huffenberger}, K.~M. and {Hurier}, G. and {Jaffe}, A.~H. and {Jaffe}, T.~R. and {Jones}, W.~C. and {Juvela}, M. and {Keih{\"a}nen}, E. and {Keskitalo}, R. and {Kisner}, T.~S. and {Knoche}, J. and {Kunz}, M. and {Kurki-Suonio}, H. and {Lagache}, G. and {Lamarre}, J.-M. and {Lasenby}, A. and {Lattanzi}, M. and {Lawrence}, C.~R. and {Leonardi}, R. and {Levrier}, F. and {Liguori}, M. and {Lilje}, P.~B. and {Linden-V{\o}rnle}, M. and {L{\'o}pez-Caniego}, M. and {Lubin}, P.~M. and {Mac{\'\i}as-P{\'e}rez}, J.~F. and {Maino}, D. and {Mandolesi}, N. and {Mangilli}, A. and {Maris}, M. and {Martin}, P.~G. and {Mart{\'\i}nez-Gonz{\'a}lez}, E. and {Masi}, S. and {Matarrese}, S. and {Melchiorri}, A. and {Mendes}, L. and {Mennella}, A. and {Migliaccio}, M. and {Miville-Desch{\^e}nes}, M.-A. and {Moneti}, A. and {Montier}, L. and {Morgante}, G. and {Mortlock}, D. and {Munshi}, D. and {Murphy}, J.~A. and {Naselsky}, P. and {Nati}, F. and {Netterfield}, C.~B. and {Noviello}, F. and {Novikov}, D. and {Novikov}, I. and {Oppermann}, N. and {Oxborrow}, C.~A. and {Pagano}, L. and {Pajot}, F. and {Paladini}, R. and {Paoletti}, D. and {Pasian}, F. and {Perotto}, L. and {Pettorino}, V. and {Piacentini}, F. and {Piat}, M. and {Pierpaoli}, E. and {Pietrobon}, D. and {Plaszczynski}, S. and {Pointecouteau}, E. and {Polenta}, G. and {Ponthieu}, N. and {Pratt}, G.~W. and {Prunet}, S. and {Puget}, J.-L. and {Rachen}, J.~P. and {Reinecke}, M. and {Remazeilles}, M. and {Renault}, C. and {Renzi}, A. and {Ristorcelli}, I. and {Rocha}, G. and {Rossetti}, M. and {Roudier}, G. and {Rubi{\~n}o-Mart{\'\i}n}, J.~A. and {Rusholme}, B. and {Sandri}, M. and {Santos}, D. and {Savelainen}, M. and {Savini}, G. and {Scott}, D. and {Soler}, J.~D. and {Stolyarov}, V. and {Sudiwala}, R. and {Sutton}, D. and {Suur-Uski}, A.-S. and {Sygnet}, J.-F. and {Tauber}, J.~A. and {Terenzi}, L. and {Toffolatti}, L. and {Tomasi}, M. and {Tristram}, M. and {Tucci}, M. and {Umana}, G. and {Valenziano}, L. and {Valiviita}, J. and {Van Tent}, B. and {Vielva}, P. and {Villa}, F. and {Wade}, L.~A. and {Wandelt}, B.~D. and {Wehus}, I.~K. and {Ysard}, N. and {Yvon}, D. and {Zonca}, A.},
        title = "{Planck intermediate results. XXXV. Probing the role of the magnetic field in the formation of structure in molecular clouds}",
      journal = {\aap},
         year = 2016,
        month = feb,
       volume = {586},
          eid = {A138},
        pages = {A138},
          doi = {10.1051/0004-6361/201525896},
archivePrefix = {arXiv},
       eprint = {1502.04123},
 primaryClass = {astro-ph.GA},
       adsurl = {https://ui.adsabs.harvard.edu/abs/2016A&A...586A.138P}
}

@INPROCEEDINGS{andre2014,
       author = {{Andr{\'e}}, P. and {Di Francesco}, J. and {Ward-Thompson}, D. and {Inutsuka}, S.-I. and {Pudritz}, R.~E. and {Pineda}, J.~E.},
        title = "{From Filamentary Networks to Dense Cores in Molecular Clouds: Toward a New Paradigm for Star Formation}",
    booktitle = {Protostars and Planets VI},
         year = 2014,
       editor = {{Beuther}, Henrik and {Klessen}, Ralf S. and {Dullemond}, Cornelis P. and {Henning}, Thomas},
        month = jan,
        pages = {27-51},
          doi = {10.2458/azu_uapress_9780816531240-ch002},
archivePrefix = {arXiv},
       eprint = {1312.6232},
 primaryClass = {astro-ph.GA},
       adsurl = {https://ui.adsabs.harvard.edu/abs/2014prpl.conf...27A}
}

@INPROCEEDINGS{difrancesco2007,
       author = {{di Francesco}, J. and {Evans}, II, N.~J. and {Caselli}, P. and {Myers}, P.~C. and {Shirley}, Y. and {Aikawa}, Y. and {Tafalla}, M.},
        title = "{An Observational Perspective of Low-Mass Dense Cores I: Internal Physical and Chemical Properties}",
    booktitle = {Protostars and Planets V},
         year = 2007,
       editor = {{Reipurth}, Bo and {Jewitt}, David and {Keil}, Klaus},
        month = jan,
        pages = {17},
          doi = {10.48550/arXiv.astro-ph/0602379},
archivePrefix = {arXiv},
       eprint = {astro-ph/0602379},
 primaryClass = {astro-ph},
       adsurl = {https://ui.adsabs.harvard.edu/abs/2007prpl.conf...17D}
}

@ARTICLE{arzoumanian2011,
       author = {{Arzoumanian}, D. and {Andr{\'e}}, Ph. and {Didelon}, P. and {K{\"o}nyves}, V. and {Schneider}, N. and {Men'shchikov}, A. and {Sousbie}, T. and {Zavagno}, A. and {Bontemps}, S. and {di Francesco}, J. and {Griffin}, M. and {Hennemann}, M. and {Hill}, T. and {Kirk}, J. and {Martin}, P. and {Minier}, V. and {Molinari}, S. and {Motte}, F. and {Peretto}, N. and {Pezzuto}, S. and {Spinoglio}, L. and {Ward-Thompson}, D. and {White}, G. and {Wilson}, C.~D.},
        title = "{Characterizing interstellar filaments with Herschel in IC 5146}",
      journal = {\aap},
         year = 2011,
        month = may,
       volume = {529},
          eid = {L6},
        pages = {L6},
          doi = {10.1051/0004-6361/201116596},
archivePrefix = {arXiv},
       eprint = {1103.0201},
 primaryClass = {astro-ph.GA},
       adsurl = {https://ui.adsabs.harvard.edu/abs/2011A&A...529L...6A}
}

@INPROCEEDINGS{dobbs2014,
       author = {{Dobbs}, C.~L. and {Krumholz}, M.~R. and {Ballesteros-Paredes}, J. and {Bolatto}, A.~D. and {Fukui}, Y. and {Heyer}, M. and {Mac Low}, Mordecai-Mark and {Ostriker}, E.~C. and {V{\'a}zquez-Semadeni}, E.},
        title = "{Formation of Molecular Clouds and Global Conditions for Star Formation}",
    booktitle = {Protostars and Planets VI},
         year = 2014,
       editor = {{Beuther}, Henrik and {Klessen}, Ralf S. and {Dullemond}, Cornelis P. and {Henning}, Thomas},
        month = jan,
        pages = {3-26},
          doi = {10.2458/azu_uapress_9780816531240-ch001},
archivePrefix = {arXiv},
       eprint = {1312.3223},
 primaryClass = {astro-ph.GA},
       adsurl = {https://ui.adsabs.harvard.edu/abs/2014prpl.conf....3D}
}

@ARTICLE{pokhrel2018,
       author = {{Pokhrel}, Riwaj and {Myers}, Philip C. and {Dunham}, Michael M. and {Stephens}, Ian W. and {Sadavoy}, Sarah I. and {Zhang}, Qizhou and {Bourke}, Tyler L. and {Tobin}, John J. and {Lee}, Katherine I. and {Gutermuth}, Robert A. and {Offner}, Stella S.~R.},
        title = "{Hierarchical Fragmentation in the Perseus Molecular Cloud: From the Cloud Scale to Protostellar Objects}",
      journal = {\apj},
         year = 2018,
        month = jan,
       volume = {853},
       number = {1},
          eid = {5},
        pages = {5},
          doi = {10.3847/1538-4357/aaa240},
archivePrefix = {arXiv},
       eprint = {1712.04960},
 primaryClass = {astro-ph.GA},
       adsurl = {https://ui.adsabs.harvard.edu/abs/2018ApJ...853....5P}
}

@ARTICLE{ebisawa2020,
       author = {{Ebisawa}, Yuji and {Sakai}, Nami and {Menten}, Karl M. and {Oya}, Yoko and {Yamamoto}, Satoshi},
        title = "{Temperature Structure of the Pipe Nebula Studied by the Intensity Anomaly of the OH 18 cm Transition}",
      journal = {\apj},
         year = 2020,
        month = dec,
       volume = {904},
       number = {2},
          eid = {136},
        pages = {136},
          doi = {10.3847/1538-4357/abc16f},
archivePrefix = {arXiv},
       eprint = {2010.06977},
 primaryClass = {astro-ph.GA},
       adsurl = {https://ui.adsabs.harvard.edu/abs/2020ApJ...904..136E}
}

@ARTICLE{harju2000,
       author = {{Harju}, J. and {Winnberg}, A. and {Wouterloot}, J.~G.~A.},
        title = "{The distribution of OH in Taurus Molecular Cloud-1}",
      journal = {\aap},
         year = 2000,
        month = jan,
       volume = {353},
        pages = {1065-1073},
       adsurl = {https://ui.adsabs.harvard.edu/abs/2000A&A...353.1065H}
}

@INPROCEEDINGS{heiles1991,
       author = {{Heiles}, C. and {Goodman}, A.~A. and {McKee}, C.~F. and {Zweibel}, E.~G.},
        title = "{Magnetic Fields in Dense Regions}",
    booktitle = {Fragmentation of Molecular Clouds and Star Formation},
         year = 1991,
       editor = {{Falgarone}, Edith and {Boulanger}, F. and {Duvert}, G.},
       series = {IAU Symposium},
       volume = {147},
        month = jan,
        pages = {43},
       adsurl = {https://ui.adsabs.harvard.edu/abs/1991IAUS..147...43H}
}

@ARTICLE{girart1999,
       author = {{Girart}, Jos{\'e} M. and {Crutcher}, Richard M. and {Rao}, Ramprasad},
        title = "{Detection of Polarized CO Emission from the Molecular Outflow in NGC 1333 IRAS 4A}",
      journal = {\apjl},
         year = 1999,
        month = nov,
       volume = {525},
       number = {2},
        pages = {L109-L112},
          doi = {10.1086/312345},
       adsurl = {https://ui.adsabs.harvard.edu/abs/1999ApJ...525L.109G}
}

@ARTICLE{ebisawa2019,
       author = {{Ebisawa}, Yuji and {Sakai}, Nami and {Menten}, Karl M. and {Yamamoto}, Satoshi},
        title = "{The Effect of Far-infrared Radiation on the Hyperfine Anomaly of the OH 18 cm Transition}",
      journal = {\apj},
         year = 2019,
        month = jan,
       volume = {871},
       number = {1},
          eid = {89},
        pages = {89},
          doi = {10.3847/1538-4357/aaf72b},
archivePrefix = {arXiv},
       eprint = {1901.10157},
 primaryClass = {astro-ph.GA},
       adsurl = {https://ui.adsabs.harvard.edu/abs/2019ApJ...871...89E}
}

@ARTICLE{poidevin2013,
       author = {{Poidevin}, Fr{\'e}d{\'e}rick and {Falceta-Gon{\c{c}}alves}, Diego and {Kowal}, Grzegorz and {de Gouveia Dal Pino}, Elisabete and {M{\'a}rio Magalh{\~a}es}, Antonio},
        title = "{Magnetic Field Components Analysis of the SCUPOL 850 {\ensuremath{\mu}}m Polarization Data Catalog}",
      journal = {\apj},
         year = 2013,
        month = nov,
       volume = {777},
       number = {2},
          eid = {112},
        pages = {112},
          doi = {10.1088/0004-637X/777/2/112},
       adsurl = {https://ui.adsabs.harvard.edu/abs/2013ApJ...777..112P}
}

@ARTICLE{roberts1995,
       author = {{Roberts}, D.~A. and {Crutcher}, R.~M. and {Troland}, T.~H.},
        title = "{The Distribution of Molecular and Neutral Gas and Magnetic Fields near the Bipolar H II Region S106}",
      journal = {\apj},
         year = 1995,
        month = mar,
       volume = {442},
        pages = {208},
          doi = {10.1086/175436},
       adsurl = {https://ui.adsabs.harvard.edu/abs/1995ApJ...442..208R}
}

@PHDTHESIS{robishaw2008,
       author = {{Robishaw}, Timothy},
        title = "{Magnetic fields near and far: Galactic and extragalactic single-dish radio observations of the Zeeman effect}",
       school = {University of California, Berkeley},
         year = 2008,
        month = dec,
       adsurl = {https://ui.adsabs.harvard.edu/abs/2008PhDT........13R}
}

@INPROCEEDINGS{robishaw2021,
       author = {{Robishaw}, Timothy and {Heiles}, Carl},
        title = "{The Measurement of Polarization in Radio Astronomy}",
    booktitle = {The WSPC Handbook of Astronomical Instrumentation, Volume 1: Radio Astronomical Instrumentation},
         year = 2021,
       editor = {{Wolszczan}, Alex},
        pages = {127-158},
          doi = {10.1142/9789811203770_0006},
       adsurl = {https://ui.adsabs.harvard.edu/abs/2021hai1.book..127R}
}

@ARTICLE{sarma2000,
       author = {{Sarma}, A.~P. and {Troland}, T.~H. and {Roberts}, D.~A. and {Crutcher}, R.~M.},
        title = "{VLA OH and H I Zeeman Observations of the NGC 6334 Complex}",
      journal = {\apj},
         year = 2000,
        month = apr,
       volume = {533},
       number = {1},
        pages = {271-280},
          doi = {10.1086/308667},
archivePrefix = {arXiv},
       eprint = {astro-ph/9912197},
 primaryClass = {astro-ph},
       adsurl = {https://ui.adsabs.harvard.edu/abs/2000ApJ...533..271S}
}

@ARTICLE{sarma2001,
       author = {{Sarma}, A.~P. and {Troland}, T.~H. and {Romney}, J.~D.},
        title = "{Very Long Baseline Array Observations of the Zeeman Effect in H$_{2}$O Masers in W3 IRS 5}",
      journal = {\apjl},
         year = 2001,
        month = jun,
       volume = {554},
       number = {2},
        pages = {L217-L220},
          doi = {10.1086/321705},
       adsurl = {https://ui.adsabs.harvard.edu/abs/2001ApJ...554L.217S}
}

@ARTICLE{sarma2002,
       author = {{Sarma}, A.~P. and {Troland}, T.~H. and {Crutcher}, R.~M. and {Roberts}, D.~A.},
        title = "{Magnetic Fields in Shocked Regions: Very Large Array Observations of H$_{2}$O Masers}",
      journal = {\apj},
         year = 2002,
        month = dec,
       volume = {580},
       number = {2},
        pages = {928-937},
          doi = {10.1086/343799},
       adsurl = {https://ui.adsabs.harvard.edu/abs/2002ApJ...580..928S}
}

@ARTICLE{sarma2013,
       author = {{Sarma}, A.~P. and {Brogan}, C.~L. and {Bourke}, T.~L. and {Eftimova}, M. and {Troland}, T.~H.},
        title = "{Very Large Array OH Zeeman Observations of the Star-forming Region S88B}",
      journal = {\apj},
         year = 2013,
        month = apr,
       volume = {767},
       number = {1},
          eid = {24},
        pages = {24},
          doi = {10.1088/0004-637X/767/1/24},
archivePrefix = {arXiv},
       eprint = {1302.6098},
 primaryClass = {astro-ph.GA},
       adsurl = {https://ui.adsabs.harvard.edu/abs/2013ApJ...767...24S}
}

@ARTICLE{sarma2009,
       author = {{Sarma}, A.~P. and {Momjian}, E.},
        title = "{Detection of the Zeeman Effect in the 36 GHz Class I CH$_{3}$OH Maser Line with the EVLA}",
      journal = {\apjl},
         year = 2009,
        month = nov,
       volume = {705},
       number = {2},
        pages = {L176-L179},
          doi = {10.1088/0004-637X/705/2/L176},
archivePrefix = {arXiv},
       eprint = {0910.1081},
 primaryClass = {astro-ph.GA},
       adsurl = {https://ui.adsabs.harvard.edu/abs/2009ApJ...705L.176S}
}

@ARTICLE{sarma2011,
       author = {{Sarma}, A.~P. and {Momjian}, E.},
        title = "{Discovery of the Zeeman Effect in the 44 GHz Class I Methanol (CH$_{3}$OH) Maser Line}",
      journal = {\apjl},
         year = 2011,
        month = mar,
       volume = {730},
       number = {1},
          eid = {L5},
        pages = {L5},
          doi = {10.1088/2041-8205/730/1/L5},
archivePrefix = {arXiv},
       eprint = {1102.2411},
 primaryClass = {astro-ph.GA},
       adsurl = {https://ui.adsabs.harvard.edu/abs/2011ApJ...730L...5S}
}

@ARTICLE{sarma2020,
       author = {{Sarma}, A.~P. and {Momjian}, E.},
        title = "{A Curious Case of Circular Polarization in the 25 GHz Methanol Maser Line toward OMC-1}",
      journal = {\apj},
         year = 2020,
        month = feb,
       volume = {890},
       number = {1},
          eid = {6},
        pages = {6},
          doi = {10.3847/1538-4357/ab6218},
archivePrefix = {arXiv},
       eprint = {1912.06235},
 primaryClass = {astro-ph.GA},
       adsurl = {https://ui.adsabs.harvard.edu/abs/2020ApJ...890....6S}
}

@ARTICLE{silverglate1984,
       author = {{Silverglate}, P.~R.},
        title = "{Upper limits to magnetic fields in CII regions.}",
      journal = {\apj},
         year = 1984,
        month = apr,
       volume = {279},
        pages = {694-697},
          doi = {10.1086/161933},
       adsurl = {https://ui.adsabs.harvard.edu/abs/1984ApJ...279..694S}
}

@ARTICLE{seo2019,
       author = {{Seo}, Young Min and {Majumdar}, Liton and {Goldsmith}, Paul F. and {Shirley}, Yancy L. and {Willacy}, Karen and {Ward-Thompson}, Derek and {Friesen}, Rachel and {Frayer}, David and {Church}, Sarah E. and {Chung}, Dongwoo and {Cleary}, Kieran and {Cunningham}, Nichol and {Devaraj}, Kiruthika and {Egan}, Dennis and {Gaier}, Todd and {Gawande}, Rohit and {Gundersen}, Joshua O. and {Harris}, Andrew I. and {Kangaslahti}, Pekka and {Readhead}, Anthony C.~S. and {Samoska}, Lorene and {Sieth}, Matthew and {Stennes}, Michael and {Voll}, Patricia and {White}, Steve},
        title = "{An Ammonia Spectral Map of the L1495-B218 Filaments in the Taurus Molecular Cloud. II. CCS and HC$_{7}$N Chemistry and Three Modes of Star Formation in the Filaments}",
      journal = {\apj},
         year = 2019,
        month = feb,
       volume = {871},
       number = {2},
          eid = {134},
        pages = {134},
          doi = {10.3847/1538-4357/aaf887},
archivePrefix = {arXiv},
       eprint = {1812.06121},
 primaryClass = {astro-ph.GA},
       adsurl = {https://ui.adsabs.harvard.edu/abs/2019ApJ...871..134S}
}

@INPROCEEDINGS{shinnaga1999,
       author = {{Shinnaga}, H. and {Tsuboi}, M. and {Kasuga}, T.},
        title = "{CCS Zeeman Observation toward a Dense Molecular Core in the Taurus Molecular Cloud}",
    booktitle = {Star Formation 1999},
         year = 1999,
       editor = {{Nakamoto}, T.},
        month = dec,
        pages = {175-176},
       adsurl = {https://ui.adsabs.harvard.edu/abs/1999sf99.proc..175S}
}

@ARTICLE{shinnaga2000,
       author = {{Shinnaga}, Hiroko and {Yamamoto}, Satoshi},
        title = "{Zeeman Effect on the Rotational Levels of CCS and SO in the $^{3}${\ensuremath{\Sigma}}$^{-}$ Ground State}",
      journal = {\apj},
         year = 2000,
        month = nov,
       volume = {544},
       number = {1},
        pages = {330-335},
          doi = {10.1086/317212},
       adsurl = {https://ui.adsabs.harvard.edu/abs/2000ApJ...544..330S}
}

@ARTICLE{surcis2009,
       author = {{Surcis}, G. and {Vlemmings}, W.~H.~T. and {Dodson}, R. and {van Langevelde}, H.~J.},
        title = "{Methanol masers probing the ordered magnetic field of W75N}",
      journal = {\aap},
         year = 2009,
        month = nov,
       volume = {506},
       number = {2},
        pages = {757-761},
          doi = {10.1051/0004-6361/200912790},
archivePrefix = {arXiv},
       eprint = {0908.3585},
 primaryClass = {astro-ph.SR},
       adsurl = {https://ui.adsabs.harvard.edu/abs/2009A&A...506..757S}
}

@ARTICLE{surcis2015,
       author = {{Surcis}, G. and {Vlemmings}, W.~H.~T. and {van Langevelde}, H.~J. and {Hutawarakorn Kramer}, B. and {Bartkiewicz}, A. and {Blasi}, M.~G.},
        title = "{EVN observations of 6.7 GHz methanol maser polarization in massive star-forming regions. III. The flux-limited sample}",
      journal = {\aap},
         year = 2015,
        month = jun,
       volume = {578},
          eid = {A102},
        pages = {A102},
          doi = {10.1051/0004-6361/201425420},
archivePrefix = {arXiv},
       eprint = {1504.06325},
 primaryClass = {astro-ph.SR},
       adsurl = {https://ui.adsabs.harvard.edu/abs/2015A&A...578A.102S}
}

@ARTICLE{surcis2022,
       author = {{Surcis}, G. and {Vlemmings}, W.~H.~T. and {van Langevelde}, H.~J. and {Hutawarakorn Kramer}, B. and {Bartkiewicz}, A.},
        title = "{EVN observations of 6.7 GHz methanol maser polarization in massive star-forming regions. V. Completion of the flux-limited sample}",
      journal = {\aap},
         year = 2022,
        month = feb,
       volume = {658},
          eid = {A78},
        pages = {A78},
          doi = {10.1051/0004-6361/202142125},
archivePrefix = {arXiv},
       eprint = {2111.08023},
 primaryClass = {astro-ph.GA},
       adsurl = {https://ui.adsabs.harvard.edu/abs/2022A&A...658A..78S}
}

@ARTICLE{suzuki1992,
       author = {{Suzuki}, Hiroko and {Yamamoto}, Satoshi and {Ohishi}, Masatoshi and {Kaifu}, Norio and {Ishikawa}, Shin-Ichi and {Hirahara}, Yasuhiro and {Takano}, Shuro},
        title = "{A Survey of CCS, HC 3N, HC 5N, and NH 3 toward Dark Cloud Cores and Their Production Chemistry}",
      journal = {\apj},
         year = 1992,
        month = jun,
       volume = {392},
        pages = {551},
          doi = {10.1086/171456},
       adsurl = {https://ui.adsabs.harvard.edu/abs/1992ApJ...392..551S}
}

@ARTICLE{tang2024,
       author = {{Tang}, Mengyao and {Qin}, Sheng-Li and {Liu}, Tie and {Zapata}, Luis A. and {Liu}, Xunchuan and {Peng}, Yaping and {Xu}, Fengwei and {Zhang}, Chao and {Tatematsu}, Ken'ichi},
        title = "{A Survey of Sulfur-bearing Molecular Lines toward the Dense Cores in 11 Massive Protoclusters}",
      journal = {\apjs},
         year = 2024,
        month = dec,
       volume = {275},
       number = {2},
          eid = {25},
        pages = {25},
          doi = {10.3847/1538-4365/ad7df0},
archivePrefix = {arXiv},
       eprint = {2409.13231},
 primaryClass = {astro-ph.GA},
       adsurl = {https://ui.adsabs.harvard.edu/abs/2024ApJS..275...25T}
}

@INPROCEEDINGS{thompson2015,
       author = {{Thompson}, M. and {Beuther}, H. and {Dickinson}, C. and {MOttram}, J. and {Klaassen}, P. and {Ginsburg}, A. and {Longmore}, S. and {Remijan}, A. and {Menten}, K.~M.},
        title = "{The ionised,radical and molecular Milky Way: spectroscopic surveys with the SKA}",
    booktitle = {Advancing Astrophysics with the Square Kilometre Array (AASKA14)},
         year = 2015,
        month = apr,
          eid = {126},
        pages = {126},
          doi = {10.22323/1.215.0126},
archivePrefix = {arXiv},
       eprint = {1412.5554},
 primaryClass = {astro-ph.GA},
       adsurl = {https://ui.adsabs.harvard.edu/abs/2015aska.confE.126T}
}

@ARTICLE{thum1999,
       author = {{Thum}, C. and {Morris}, D.},
        title = "{A strong magnetic field in the disk of MWC 349}",
      journal = {\aap},
         year = 1999,
        month = apr,
       volume = {344},
        pages = {923-929},
       adsurl = {https://ui.adsabs.harvard.edu/abs/1999A&A...344..923T}
}

@ARTICLE{troland1982,
       author = {{Troland}, T.~H. and {Heiles}, C.},
        title = "{The Zeeman effect in 21 centimeter line radiation - Methods and initial results}",
      journal = {\apj},
         year = 1982,
        month = jan,
       volume = {252},
        pages = {179-192},
          doi = {10.1086/159544},
       adsurl = {https://ui.adsabs.harvard.edu/abs/1982ApJ...252..179T}
}

@ARTICLE{thompson2019,
       author = {{Thompson}, K.~L. and {Troland}, T.~H. and {Heiles}, C.},
        title = "{A Survey of Magnetic Field Strengths in the Envelopes of Molecular Clouds via the 18 cm OH Zeeman Effect}",
      journal = {\apj},
         year = 2019,
        month = oct,
       volume = {884},
       number = {1},
          eid = {49},
        pages = {49},
          doi = {10.3847/1538-4357/ab364e},
archivePrefix = {arXiv},
       eprint = {1907.11940},
 primaryClass = {astro-ph.GA},
       adsurl = {https://ui.adsabs.harvard.edu/abs/2019ApJ...884...49T}
}

@ARTICLE{troland2008,
       author = {{Troland}, Thomas H. and {Crutcher}, Richard M.},
        title = "{Magnetic Fields in Dark Cloud Cores: Arecibo OH Zeeman Observations}",
      journal = {\apj},
         year = 2008,
        month = jun,
       volume = {680},
       number = {1},
        pages = {457-465},
          doi = {10.1086/587546},
archivePrefix = {arXiv},
       eprint = {0802.2253},
 primaryClass = {astro-ph},
       adsurl = {https://ui.adsabs.harvard.edu/abs/2008ApJ...680..457T}
}

@ARTICLE{turner2006,
       author = {{Turner}, B.~E. and {Heiles}, Carl},
        title = "{The C$_{4}$H Zeeman Effect in TMC-1: Understanding Low-Mass Star Formation}",
      journal = {\apjs},
         year = 2006,
        month = feb,
       volume = {162},
       number = {2},
        pages = {388-400},
          doi = {10.1086/498431},
       adsurl = {https://ui.adsabs.harvard.edu/abs/2006ApJS..162..388T}
}

@ARTICLE{vlemmings2001,
       author = {{Vlemmings}, W. and {Diamond}, P.~J. and {van Langevelde}, H.~J.},
        title = "{Circular polarization of circumstellar water masers around S Per}",
      journal = {\aap},
         year = 2001,
        month = aug,
       volume = {375},
        pages = {L1-L4},
          doi = {10.1051/0004-6361:20010890},
archivePrefix = {arXiv},
       eprint = {astro-ph/0107070},
 primaryClass = {astro-ph},
       adsurl = {https://ui.adsabs.harvard.edu/abs/2001A&A...375L...1V}
}

@ARTICLE{vlemmings2005,
       author = {{Vlemmings}, W.~H.~T. and {van Langevelde}, H.~J. and {Diamond}, P.~J.},
        title = "{The magnetic field around late-type stars revealed by the circumstellar H\_2O masers}",
      journal = {\aap},
         year = 2005,
        month = may,
       volume = {434},
       number = {3},
        pages = {1029-1038},
          doi = {10.1051/0004-6361:20042488},
archivePrefix = {arXiv},
       eprint = {astro-ph/0501628},
 primaryClass = {astro-ph},
       adsurl = {https://ui.adsabs.harvard.edu/abs/2005A&A...434.1029V}
}
